\documentclass[
 reprint,
 amsmath,amssymb,
 aps,
 prx,
 longbibliography
]{revtex4-2}

\usepackage{natbib}
\usepackage{graphicx}
\usepackage{dcolumn}
\usepackage{bm}
\usepackage{float}
\usepackage{placeins}
\usepackage{listings}
\usepackage{xcolor}
\usepackage{braket}
\usepackage{amsmath}
\usepackage{MnSymbol}
\usepackage{svg}
\usepackage{textcomp}
\usepackage{mathtools}
\usepackage[normalem]{ulem}
\usepackage{hyperref}
\usepackage{silence}
\usepackage{booktabs}
\usepackage{multirow}
\usepackage[normalem]{ulem}
\begin{document}

\preprint{APS/123-QED}

\title{Thermometry Based on a Superconducting Qubit}

\author{D. S. Lvov}
 \email{dmitrii.lvov@aalto.fi}

\author{S. A. Lemziakov}

\author{E. Ankerhold}

\author{J. T. Peltonen}

\author{J. P. Pekola}
 
\affiliation{PICO group, QTF Centre of Excellence, Department of Applied Physics, Aalto University, P.O. Box 15100, FI-00076 Aalto, Finland}%

\date{\today}

\begin{abstract}
We report temperature measurements using a transmon qubit by detecting the population of its first three energy levels, after applying a sequence of $\pi$-pulses and performing projective dispersive readout. We measure the effective temperature of the qubit and characterize its relaxation and coherence times $\tau_{1,2}$ for three devices in the temperature range of $20-300$\,mK. We analyze the process of qubit thermalization to its effective environment consisting of multiple heat baths and support it with experimental data. Signal-to-noise (SNR) ratio of the temperature measurement depends strongly on $\tau_1$, which drops at higher temperatures due to quasiparticle excitations, adversely affecting the measurements and setting an upper bound of the dynamic temperature range of the thermometer. The measurement relies on coherent dynamics of the qubit during the $\pi$-pulses. The effective qubit temperature follows closely that of the cryostat in the range of $100 - 250$\,mK. We present a numerical model of the qubit population distribution and compare it favorably with the experimental results. Finally, we compare our technique with previous works on qubit thermometry and discuss its application prospects.
\end{abstract}

\maketitle

\section{\label{sec:intro}Introduction}

In recent years researchers demonstrated experiments on thermometry of various quantum systems, such as cold gases \cite{Bouton_2020, Mitchison_2020, Glatthard_2022}, ion crystals \cite{Vybornyi_2023}, spin ensembles in solutions \cite{Raitz_2015, Uhlig_2019}, NV-centers \cite{Kucsko_2013, Neumann_2013, Toyli_2013, Pingault_2017, Fujiwara_2021}, semiconductor quantum dots \cite{Torresani_2013, Ahmed_2018, Nicolí_2019, Champain_2024} and superconducting qubits \cite{Geerlings_2013, Jin_2015, Kulikov_2020, Scigliuzzo_2020, Sultanov_2021, Sharafiev_2024}. Quantum thermometry of coherent systems attracts special attention due to prospects of monitoring temperature with minimal invasion, properties of quantum systems with high accuracy at extremely low temperatures. It also provides opportunities for studying open quantum systems and quantum thermodynamics \cite{Brunelli_2011, Higgins_2013, Jevtic_2015, Correa_2017, Braun_2018, Mehboudi_2019, Mukherjee_2019, Gebbia_2020, Rubio_2021, Mitchison_2020, Seah_2019, Jorgensen_2020, Latune_2020, Román-Ancheyta_2020, Pati_2020, Khan_2022, Zhang_2022, Glatthard_2022, Zakharov_2024, Sharafiev_2024}.

Superconducting qubits are a well-established platform for realization of quantum electrodynamics (QED) \cite{Blais_2021} which is widely used for building multi-qubit systems \cite{Blais_2004, Arute_2019, Krantz_2019, McEwen_2022}. 
However, measurements of the population distribution of such qubits are mainly aimed at device characterization and benchmarking of control pulses by analyzing the residual population. 
This has been achieved using measurements of Rabi oscillations \cite{Geerlings_2013, Jin_2015}, applying different sequences of $\pi$-pulses \cite{Sultanov_2021}, correlation measurements \cite{Kulikov_2020} or single-shot measurements (see for example Ref.\,\cite{Teixeira_2024}).
Another approach was demonstrated on the basis of temperature dependent S-parameter of a microwave waveguide directly coupled to a qubit \cite{Scigliuzzo_2020} or a two qubit system \cite{Sharafiev_2024}.

Qubits are sensitive to various sources of dissipation and decoherence, and the diverse nature of their environment (see, for example, \cite{Schoelkopf_2003, Krantz_2019, Catelani_2012, Lisenfeld_2019, Abdurakhimov_2022, Iaia_2022, Pan_2022, Thorbeck_2023, Kono_2023}). Therefore, in practice, identifying which specific thermal bath leads to thermalization of the qubit, can be quite challenging; i.e., it may be difficult to identify what temperature the qubit is actually measuring. Thus, prospects of utilizing superconducting qubits as thermometers are of dual research interest: they can facilitate studies of open quantum systems and interaction between the qubits and their environment, which is vital in the context of building large multiqubit systems such as quantum computers and simulators. On the other hand, qubits can be used for realization of various quantum thermometry protocols. 

In this paper we present a comprehensive experimental study of employing a superconducting qubit for quantum thermometry at sub-kelvin temperatures. The measurement technique used in our work relies on applying sequences of $\pi$-pulses for manipulating the qubit population distribution, requiring sufficient coherence of the qubit, and then performing consecutive qubit state readout, which was described in Ref.\,\cite{Sultanov_2021}. We observe and discuss the process of qubit thermalization to its effective environment consisting of different thermal baths, as well as temperature dependencies of qubit relaxation and coherence times. These timescales determine the working range of the thermometer in terms of temperature and introduce constraints on the measurement pulse durations. While showing a solid thermalization of the qubit to the cryostat stage, the technique allows us to measure the population distribution at temperatures up to $200-220$\,mK, where the second excited state of the transmon qubit is already significantly populated. This exceeds the applicability range of methods based on consideration of only the lowest transmon levels. At the same time, the technique is less dependent on the signal-to-noise ratio (SNR) of the measurement setup. We compare the experimental results with a numerical model of qubit population measurements. Within the model we investigate the influence of the measured parameters on the population and effective qubit temperature obtained as the result. Finally, we address the factors limiting the dynamical range of a superconducting qubit thermometer and perspective applications. The paper is organized as follows: in Section\,\ref{sec:qubit-bath} we provide a description of a realistic model of a qubit being in thermal equilibrium with several heat baths. Section\,\ref{sec:pop_meas} explains the principle of the population distribution measurements. Section\,\ref{sec:exp_details} provides a brief description of the samples used in the experiment, measurement sequence, and discussion on the main experimental results. Finally, limits of the thermometer, ways of improvement and possible application, as well as comparison with other experimental works are discussed in Section\,\ref{sec:limits}. The paper is followed by appendices providing additional experimental data, details of the numerical simulations, experimental setup and sample fabrication.

\section{\label{sec:qubit-bath}Qubit population and effective temperature}

Let us consider a qubit in thermal equilibrium with a heat bath having temperature $T$. In this case the qubit is in a thermal state, which means that its density matrix is diagonal and the probabilities $p_g$ and $p_e$, respectively, in the ground $\ket{g}$ or excited state $\ket{e}$ (qubit populations) are Boltzmann distributed as
\begin{equation}
\label{eq:pepg}
    p_e/p_g = e^{-\frac{\hbar \omega_{ge}}{k_B T}},
\end{equation}
where $\hbar\omega_{ge}$ is the separation between these two states, and $k_B$ is the Boltzmann constant. Thus, by measuring the population ratio of the qubit one obtains the temperature of the thermal bath. 

In a more realistic scenario where a qubit is in a steady state with respect to several uncorrelated heat baths (with temperatures $T^{(i)}$, where $i$ enumerates baths), the state of the qubit is still a thermal-like state. Excitation $\left(\Gamma_\uparrow^{(i)}\right)$ and relaxation rates $\left(\Gamma_\downarrow^{(i)}\right)$ caused by each bath follow the detailed balance principle
\begin{equation}
    \label{eq:det_balance}
    \frac{\Gamma_\uparrow^{(i)}}{\Gamma_\downarrow^{(i)}} = \exp{\left(-\frac{\hbar\omega_{ge}}{k_B T^{(i)}}\right)}.
\end{equation}

However, the temperature characterizing such qubit state is not necessarily the temperature $T^{(i)}$ of any of the thermal baths. Then Eq.~(\ref{eq:pepg}) could be interpreted as giving an effective qubit temperature $T_\mathrm{eff}$. In this sense, Eq.\,(\ref{eq:pepg}) can be rewritten as a definition of $T_\mathrm{eff}$ as
\begin{equation}
    \label{eq:Teff}
    T_\mathrm{eff} = \frac{\hbar \omega_{ge}}{k_B} \left(- \ln{\frac{p_e}{p_g}}\right)^{-1},
\end{equation}
which coincides with the true temperature in equilibrium. For the total qubit excitation and relaxation rates $\Gamma_\uparrow = \Sigma_i \Gamma_\uparrow^{(i)}$ and $\Gamma_\downarrow = \Sigma_i \Gamma_\downarrow^{(i)}$ the detail balance principle is also valid:
\begin{equation}
    \label{eq:qubit_det_balance}
    \frac{\Gamma_\uparrow}{\Gamma_\downarrow}=\frac{\sum_i\Gamma_\uparrow^{(i)}}{\sum_i\Gamma_\downarrow^{(i)}} = \exp{\left(-\frac{\hbar\omega_{ge}}{k_B T_\mathrm{eff}}\right)}.
\end{equation}

Then we can introduce a photon number
\begin{equation}
    \label{eq:n_eff}
    n_\mathrm{eff} = \frac{1}{\exp{\left(\hbar\omega_{ge}/k_B T_\mathrm{eff}\right)} - 1}=\frac{\sum_i\Gamma_\uparrow^{(i)}}{\sum_i\Gamma_\downarrow^{(i)} - \Gamma_\uparrow^{(i)}}.
\end{equation}

If the baths have ohmic spectrum, the rates write
\begin{align}
\label{eq:rates_ohmic}
    &\Gamma_\uparrow^{(i)} = \gamma^{(i)} n_i\left(\omega_{ge}, T^{(i)}\right),\\
    &\Gamma_\downarrow^{(i)} = \gamma^{(i)} \left[n_i(\omega_{ge}, T^{(i)}) + 1\right],
\end{align}
where $\gamma^{(i)}$ are constant coupling rates, which could be obtained from the Fermi golden rule, and $n_i \left(\omega_{ge}, T^{(i)}\right) = \left[\exp{\left(\hbar\omega_{ge}/k_B T^{(i)}\right)} - 1\right]^{-1}$ is the Bose-Einstein distribution.
The qubit energy relaxation rate then writes
\begin{align}
\label{eq:gamma1_2baths}
    \gamma_1 = \sum_{i} \left(\Gamma_{\downarrow}^{(i)}+\Gamma_{\uparrow}^{(i)}\right)=\sum_{i} \gamma^{(i)}&\left(2n_i+1\right)\nonumber\\
    = \gamma_1^0 \left(2n_\mathrm{eff}+1\right)&=\gamma_1^0\coth{\frac{\hbar \omega_{ge}}{2 k_B T_\mathrm{eff}}},
\end{align}
where $\gamma_1^0 = \sum_{i} \gamma^{(i)}$ is the total relaxation rate at zero temperature. Finally, we can rewrite $n_\mathrm{eff}$ from Eq.\,(\ref{eq:n_eff}) as
\begin{equation}
\label{eq:n_eff_2}
    n_\mathrm{eff} = \frac{1}{\gamma_1^0}\sum_i \gamma^{(i)} n_i,
\end{equation}
which can be considered as the average photon number over different baths, weighted by their coupling coefficients.

\section{\label{sec:pop_meas}Population measurements}

\begin{figure*}[tbh!]
\includegraphics[width=1.00\textwidth]{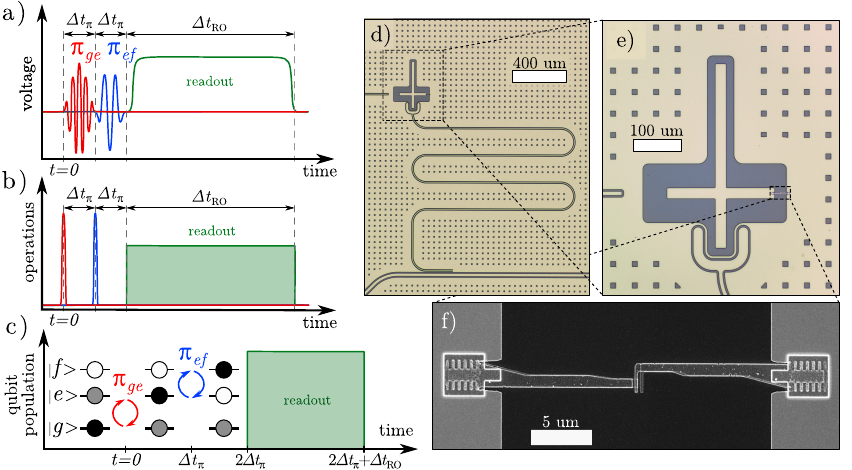}
\caption{\label{fig:chip_pulses} 
Scheme of the experiment and the sample design. 
a) The pulse sequence used in the $x_2$-measurement.
b) A scheme of the qubit operations, depicting the numerical simulation of the $x_2$-measurement.
c) A schematic depiction of how the $x_2$-sequence of $\pi$-pulses changes the qubit population distribution. Initially the qubit population corresponds to the thermalized state. Then the $\pi$-pulses consecutively swap populations of the pairs of qubit levels. Finally, there is a finite length readout duration, during which the population is still evolving. 
d) An optical image of the transmon and readout resonator coupled to the waveguide (sample Q2-III). 
e) An optical image of the transmon, shown on panel d).  
f) SEM image of the Al/AlOx/Al tunnel contact of the transmon, circled in panel e).
}
\end{figure*}

In this work we measured the population distribution of several transmon qubits to obtain their effective temperature using the technique described in Ref.\,\cite{Sultanov_2021}, which is based on applying different sequences of $\pi$-pulses and then performing a dispersive readout of the qubit state. We consider a qutrit model, which accounts for population of three lowest energy levels in the analysis and protocols. We presume, that the population of a qubit thermalized at temperature $T$ follows the Maxwell-Boltzmann distribution $p_i = \exp{\left(-E_i/k_B T\right)} /Z$, with the partition function $Z = \sum_{i=1}^{3}\exp{\left(-E_i/k_B T\right)}$ and the $i$-th level energy $E_i$.
A transmon qubit typically has a weak anharmonicity $\alpha$ $( |\alpha| \ll \hbar\omega_{ge})$ \cite{Koch_2007}, so that the energy separations $\hbar\omega_{ge}$ and $\hbar\omega_{ef}$ are related by $\omega_{ef} = \omega_{ge} + \alpha/\hbar \cong \omega_{ge}$. For example, if a transmon with $\omega_{ge}/2\pi=6.65\,$GHz and $\alpha/h = -230\,$MHz is thermalized at temperature $300\,$mK we can estimate the population ratios $p_e/p_g \approx 0.35$ and $p_f/p_g \approx 0.12$. Thus, it is important to take into account the finite population of the $f$-level. Limited to the three lowest energy states of a transmon, the density matrix of a thermalized transmon reads 
\begin{equation}
\hat{\rho}=p_g\ket{g}\bra{g}+p_e\ket{e}\bra{e}+p_f\ket{f}\bra{f}.
\end{equation}

\noindent 
Let us denote the readout response of a projective measurement of the qubit being in state $\ket{i}$ as $\varphi_i$, which we will call ``pure state response'' further in the text. Then the result $\varphi$ of the thermalized qubit state measurement is a linear combination of the form $\varphi_\mathrm{therm} = p_g \varphi_g+p_e \varphi_e+p_f \varphi_f$. 
From the experimental point of view, $\varphi$ here is a measured complex valued voltage. To measure it we send a readout pulse with a rectangular envelope at a frequency close to the readout resonator frequency $\omega_r$, which is transmitted through the sample waveguide, then it is amplified and downconverted. Finally, it is measured by a fast (2 GS/s) analog-to-digital converter (ADC) and integrated over the pulse duration time, which is typically $0.1-2.0\,\mu$s.

However, a single measurement of $\varphi_\mathrm{therm}$ is not enough to extract the populations $p_i$ as the pure state responses $\varphi_i$ are unknown. To solve this problem, before the qubit state readout is performed, $\pi_{ge}$- and $\pi_{ef}$-pulses can be applied at frequencies $\omega_{ge}$ and $\omega_{ef}$, respectively, that leads to swapping of the population between $\ket{g}$ and $\ket{e}$ states or $\ket{e}$ and $\ket{f}$ states, correspondingly. This results in different readout signals, which we denote $x_{j}$ and $y_{k}$, $j,k = 0,1,2$. An example of such a measurement is shown in Fig.\,\ref{fig:chip_pulses}\,a,\,c. For different $\pi$-pulse sequences we can then write a system of linear equations as
\begin{align}
\label{eq:xy}
    &x_0 = p_g \varphi_g + p_e \varphi_e + p_f \varphi_f,\quad\quad\left(\mathrm{no\ pulse}\right)\nonumber \\ 
    &x_1 = p_e \varphi_g + p_g \varphi_e + p_f \varphi_f,\quad\quad\quad\left(\pi_{ge}\right)\nonumber \\ 
    &x_2 = p_e \varphi_g + p_f \varphi_e + p_g \varphi_f,\quad\quad\ \left(\pi_{ge}\pi_{ef}\right)\\
    &y_0 = p_g \varphi_g + p_f \varphi_e + p_e \varphi_f,\quad\quad\quad\left(\pi_{ef}\right)\nonumber \\
    &y_1 = p_f \varphi_g + p_g \varphi_e + p_e \varphi_f,\quad\quad\ \left(\pi_{ef}\pi_{ge}\right)\nonumber \\
    &y_2 = p_f \varphi_g + p_e \varphi_e + p_g \varphi_f.\quad\quad\left(\pi_{ef}\pi_{ge}\pi_{ef}\right)\nonumber
\end{align}

\begin{table*}[t!]
\caption{\label{tab:ABC}Different ways of the qubit effective temperature calculation from the qubit state readout results.}
\begin{ruledtabular}
\begin{tabular}{cccccc}
\multirow{2}{*}{Ratio}&\multicolumn{3}{c}{Calculation method}&\multirow{2}{*}{Result}&\multirow{2}{*}{Equation}\\
&1&2&3&&\\ \hline
 
A&$\frac{x_0-x_1}{y_0-y_1}$&$\frac{y_0-x_2}{x_0-y_2}$&$\frac{y_1-y_2}{x_1-x_2}$&$\frac{p_g-p_e}{p_g-p_f}$&$A=\frac{1-\exp{\left(-\hbar\omega_{ge}/k_B T\right)}}{{1-\exp{\left(-\hbar\omega_{gf}/k_B T\right)}}}$\\
 
B&$\frac{x_2-y_2}{x_0-x_1}$&$\frac{x_1-y_1}{y_0-x_2}$&$\frac{x_0-y_0}{y_1-y_2}$&$\frac{p_e-p_f}{p_g-p_e}$&$B=\frac{\exp{\left(-\hbar\omega_{ge}/k_B T\right)}-\exp{\left(-\hbar\omega_{gf}/k_B T\right)}}{{1-\exp{\left(-\hbar\omega_{ge}/k_B T\right)}}}$\\
 
C&$\frac{x_2-y_2}{y_0-y_1}$&$\frac{x_1-y_1}{x_0-y_2}$&$\frac{x_0-y_0}{x_1-x_2}$&$\frac{p_e-p_f}{p_g-p_f}$&$C=\frac{\exp{\left(-\hbar\omega_{ge}/k_B T\right)}-\exp{\left(-\hbar\omega_{gf}/k_B T\right)}}{{1-\exp{\left(-\hbar\omega_{gf}/k_B T\right)}}}$\\

\end{tabular}
\end{ruledtabular}
\end{table*}

\begin{table*}[t!]
\caption{\label{tab:q_params}Parameters of the devices. 
}
\begin{ruledtabular}
\begin{tabular}{cccccccc}
 Device & $\omega_{ge}/2\pi$,\,GHz & $\omega_{ef}/2\pi$,\,GHz & $E_c/h$,\,MHz
& $\Delta_r/2\pi$,\,GHz & $g/2\pi$,\,MHz\\ \hline
 R2-I& 6.422 & 6.221 & 201 & 1.876 & 34 \\
 R4-I& 6.649 & 6.417 & 232 & 1.753 & 38 \\
 R3-II& 6.732 & 6.513 & 219 & 0.765 & 44 \\
 Q2-III& 7.042 & 6.835 & 207 & 2.151 & 37 \\
\end{tabular}
\end{ruledtabular}
\end{table*}

We can find functional dependencies between $p_i, i\in\{g,e,f\}$ and $x_j, y_k$ 
which are valid for arbitrary pure state responses $\varphi_i$. These quantities are temperature dependent given by $A=\frac{p_g-p_e}{p_g-p_f}$, $B=\frac{p_e-p_f}{p_g-p_e}$ and $C=\frac{p_e-p_f}{p_g-p_f}$, which can also be expressed through the readout results $x_j,\ y_k$ (see Table\,\ref{tab:ABC}).
To understand the temperature behavior of functions $A,\ B$ and $C$ better, we can take the low temperature limit $\hbar\omega_{ge}\ll k_B T$, where $p_f \rightarrow 0$. Then we 
can rewrite $A \approx 1 - p_e/p_g$, $B \approx C \approx p_e/p_g$ or 
\begin{eqnarray}
\label{eq:A}
    A \approx 1 - &\exp{\left(-\hbar\omega_{ge}/k_B T\right)}, \\
\label{eq:BC}
    B \approx C \approx &\exp{\left(-\hbar\omega_{ge}/k_B T\right)}.
\end{eqnarray}

We extract the effective qubit temperature $T_\mathrm{eff}$ by numerically solving 
equations for functions $A, B$ and $C$, shown in Tab.\,\ref{tab:ABC}.

Population ratios $A,\ B$ and $C$ can be extracted from the measurement data $x_{j},\ y_{k}$ by three different methods each, which we denote as $A_1, A_2, A_3, B_1, \dots, C_3$ (see Tab.\,\ref{tab:ABC}). Qualitatively speaking, the difference between the methods is the order of the $\pi$-pulses in the control sequence. Because of the qubit relaxation and relatively low population of the $f$-level, the resulting signals can have different signal-to-noise (SNR) ratio, giving us some guidance on how to choose the best one (see discussion in Section\,\ref{sec:limits} and Appendix\,\ref{app:errors}).

\section{\label{sec:exp_details}Experimental details}
We used transmon qubits on three different chips I, II and III (which is reflected in their names: R2-I, R4-I, R3-II, Q2-III) made in different fabrication cycles. Each transmon qubits had either fixed transition frequency $\omega_{ge}/2\pi\approx6.5\,$GHz or flux-tunable spectrum ($\omega_{ge}/2\pi = 5.5-6.8\,$GHz) and charging energy $E_c/h=210-230\,$MHz (see Fig.\,\ref{fig:chip_pulses}) and they were weakly coupled ($g/2\pi = 35-45$\,MHz) to a coplanar waveguide (CPWG) $\lambda/4$ readout resonator with the internal quality factor $Q_\mathrm{int}\sim 10^5$, external quality factor $Q_\mathrm{ext}\approx 8\cdot 10^4$ and fundamental frequency $\omega_{r}/2\pi=4.5-6.0\,$GHz for a dispersive state readout ($\left|\Delta_r\right| = \left|\omega_{ge}-\omega_r\right| \gg g$).
Experimentally found parameters of the devices are shown in Table\,\ref{tab:q_params}.
The fabrication details are given in Appendix\,\ref{app:fab}.

In the experiment we measured qubit energy relaxation and coherence times $\tau_1$ and $\tau_2$, and qubit effective temperature $T_\mathrm{eff}$, while controlling the temperature of the mixing chamber (MXC) stage of the dilution refrigerator. After a certain MXC stage temperature $T_\mathrm{MXC}$ was set, we waited for 10\,min for thermalization of the sample stage and then performed a series of measurements, consisting of five repetitions of the following measurement sequence: (i) measurement of the MXC stage temperature $T_\mathrm{MXC}$ using a standard RuOx thermometer, (ii) population distribution measurement $x_j,\ y_k$, (iii) measurement of qubit relaxation time $\tau_1$ in the qubit decay experiment, (iv) Ramsey oscillations measurement to get  ``Ramsey coherence time'' $\tau_2^R$  and $\omega_{ge}$, and (v) Hahn echo measurement with the decay time $\tau_2^E$ (``echo coherence time'') \cite{Krantz_2019, Blais_2021}. Each of these iterations took approximately 5\,min. Qubit excitation pulses were applied via an AC drive antenna, and the readout pulses were sent through a waveguide. The measurement setup is shown in the Appendix in Fig.\,\ref{fig:setup}.

\begin{figure}[htb]
\includegraphics[width=0.50\textwidth]{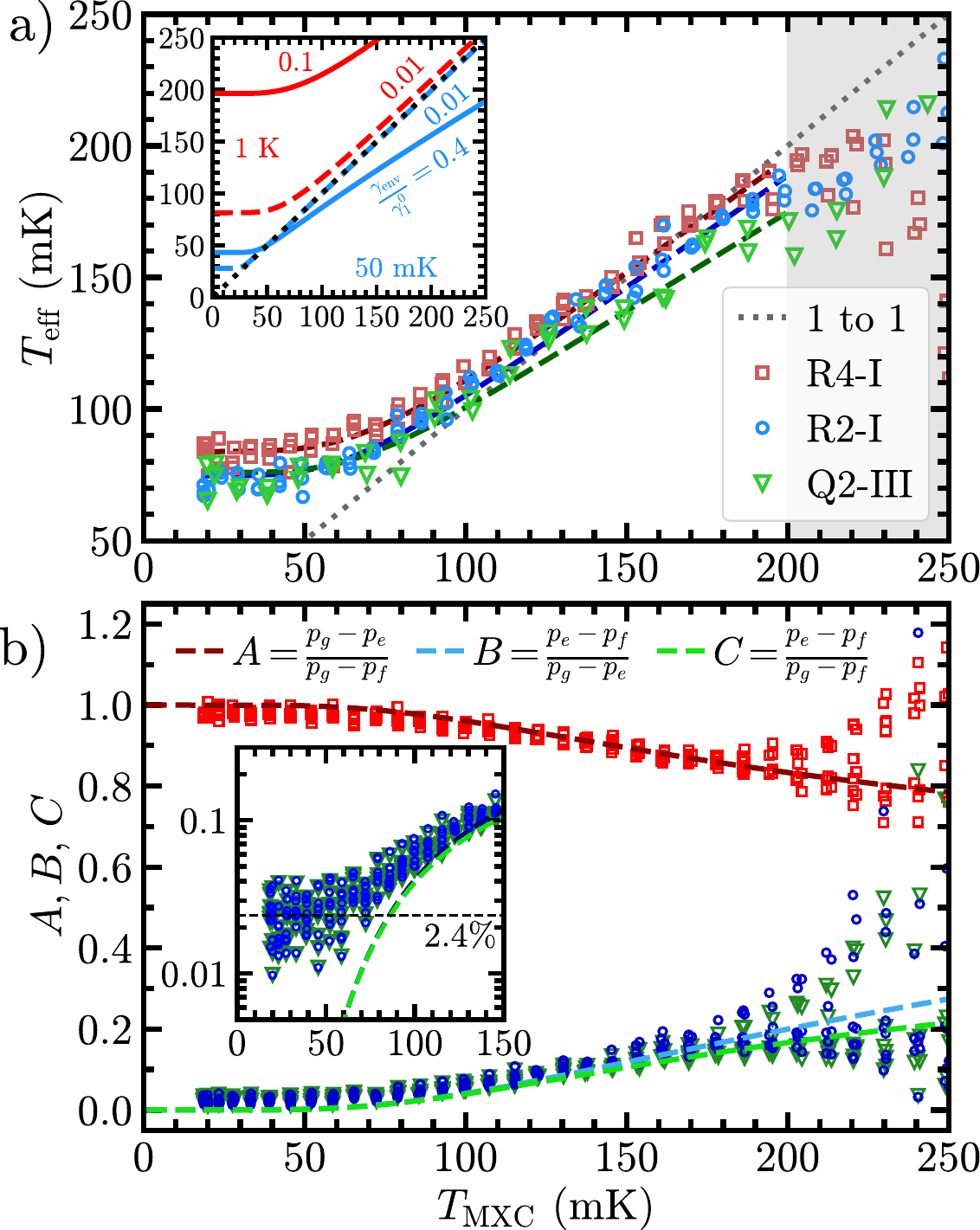}
\caption{\label{fig:exp1A}
a) Measurements of the effective qubit temperature at various temperatures of the cryostat MXC stage for different qubits extracted from the values $B_2$ (see Tab.\,\ref{tab:ABC}). The colored dashed lines show the dependence of $T_\mathrm{eff}\left(T_\mathrm{MXC}\right)$ in presence of hot effective environment with $T_\mathrm{env}$ and the gray dotted line shows the one-to-one correspondence. The vertical gray band above 200\,mK denotes the temperature range with significant quasiparticle influence on the measurements.
Inset: dependence of $T_\mathrm{eff}$ on cryostat temperature, calculated from the two-bath model (see discussion in Section\,\ref{sec:2_baths}). Blue lines show the cold bath case $\left(T_\mathrm{env}=50\,\mathrm{mK}\right)$. Blue dashed line show the regime of the weak coupling to the environment with $\gamma_\mathrm{env}/\gamma_1^0=0.01$ and the blue solid line show the strong coupling case $\gamma_\mathrm{env}/\gamma_1^0=0.4$. Red lines show the hot bath case $\left(T_\mathrm{env}=1\,\mathrm{K}\right)$. Analogously, red dashed line show the weak coupling regime with $\gamma_\mathrm{env}/\gamma_1^0=0.01$ and the red solid line show the stronger coupling case $\gamma_\mathrm{env}/\gamma_1^0=0.1$. Black dashed line shows 1-to-1 correspondence.
b) Temperature dependence of the measured population ratios $A,B$ and $C$ (green) for sample R4-I. The dashed lines show the values of $A,\ B,\ C$ according to the expressions in Tab.\,\ref{tab:ABC}. The inset shows the low temperature range for $A$ and $B$ on a logarithmic scale, showing a saturation of $p_e/p_g \approx 2.4\%$.}
\end{figure}

\begin{figure}[htb]
\includegraphics[width=0.49\textwidth]{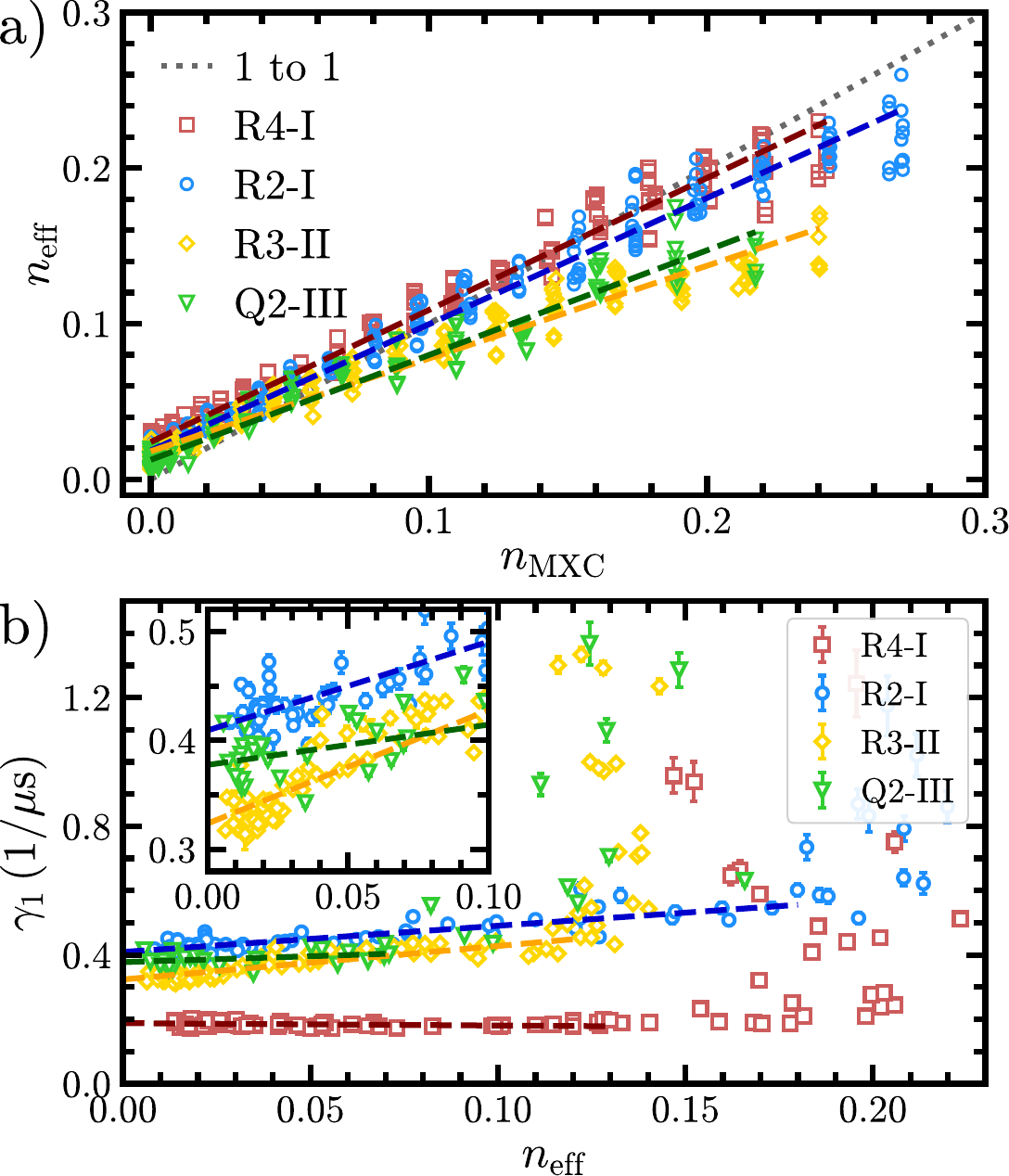}
\caption{\label{fig:exp1B}
a) Dependence between the photon numbers \\$n_\mathrm{eff} = n\left(\omega_{ge}, T_\mathrm{eff}\right)$ and $n_\mathrm{MXC} = n\left(\omega_{ge}, T_\mathrm{MXC}\right)$. The black short-dashed line shows one-to-one correspondence and the colored dashed lines are linear fits. 
b) Dependence of the relaxation rate on the photon number $n_\mathrm{eff}$. The linear region is fitted to extract the zero-temperature rate $\gamma_1^0 = \gamma_1\left(n_\mathrm{eff}=0\right)$. At larger $n_\mathrm{eff}$ one can see exponential rise of relaxation rate, caused by quasiparticles.
The fitting parameters for both plots are shown in Table\,\ref{tab:fitting_params}.
}
\end{figure}

Measurements of the qubit effective temperature and an example of population measurements in terms of datasets $A, B$ and $C$ are shown in Fig.\,\ref{fig:exp1A}. The effective qubit temperature demonstrates a linear behavior in the temperature range $100-200$\,mK close to one-to-one correspondence with the temperature of the mixing chamber stage $T_\mathrm{MXC}$ of the cryostat. At lower temperatures of the cryostat $T_\mathrm{eff}$ exhibits saturation, reaching $T_\mathrm{eff}^0\approx 65-85\,$mK, which coincides with the saturation temperature estimated from the population $p_e/p_g \approx 0.024$ in Fig.\,\ref{fig:exp1A}\,b. This elevated temperature compared to the base temperature of the cryostat can be explained by the presence of two baths, coupled to the qubit. One of these baths has the cryostat temperature $T_\mathrm{MXC}$ while the second one, to which we will refer as to ``environment'', is probably caused by imperfect shielding against thermal photons and insufficient filtering of the measurement lines, has a constant temperature $T_\mathrm{env}$. 
We discuss this model in Section\,\ref{sec:2_baths}.

Above 170\,mK the population readings in Fig.\,\ref{fig:exp1A}\,b have increased deviation from the analytical expressions caused by temperature dependent relaxation time. A numerical model, which accounts for this effect is described in Section\,\ref{sec:numerics}.

\subsection{\label{sec:2_baths}Two-bath model}
Assuming that our qubit is in thermal equilibrium, let us rewrite Eqs.\,(\ref{eq:gamma1_2baths}) and (\ref{eq:n_eff_2}) for our case of two baths with temperatures $T_\mathrm{MXC}$ and $T_\mathrm{env}$:
\begin{align}
    \label{eq:n_eff_vs_n_mxc}
    &n_\mathrm{eff} = \frac{\gamma_\mathrm{MXC}}{\gamma_1^0}\,n_\mathrm{MXC}
    +\frac{\gamma_\mathrm{env}}{\gamma_1^0}\,n_\mathrm{env},\\
    \label{eq:gamma1_vs_n_mxc}
    \gamma_1 = &\gamma_\mathrm{MXC}\,  \left(2n_\mathrm{MXC}+1\right) + \gamma_\mathrm{env}\,\left(2n_\mathrm{env}+1\right)\nonumber\\
    &=\gamma_1^0\left(2n_\mathrm{eff}+1\right),
\end{align}

\noindent where $n_\mathrm{eff} = n\left(\omega_{ge},T_\mathrm{eff}\right)$,   $n_\mathrm{MXC} = n\left(\omega_{ge},T_\mathrm{MXC}\right)$ and $n_\mathrm{env} = n\left(\omega_{ge},T_\mathrm{env}\right)$, $\gamma_1^0=\gamma_\mathrm{MXC}+\gamma_\mathrm{env}$ and coefficients $\gamma_\mathrm{MXC}$ and $\gamma_\mathrm{env}$ are temperature independent rates of relaxation to the cryostat bath and the environment, respectively. If we plot experimental data for dependence $T_\mathrm{eff}\left(T_\mathrm{MXC}\right)$ in terms of $n_\mathrm{eff}$ and $n_\mathrm{MXC}$ (see Fig.\,\ref{fig:exp1B}\,a), then we can extract from a linear fit both the ratio $\gamma_\mathrm{MXC}/\gamma_1^0$ as the slope, and the photon number $n_\mathrm{env}\left(T_\mathrm{env}\right)$ from the offset, which also gives the effective environment temperature $T_\mathrm{env} = 100-165\,$mK. Fitting results and extracted $T_\mathrm{env}$ are shown in Tab.\,\ref{tab:fitting_params}. Recalculated temperature dependence $T_\mathrm{eff}\left(T_\mathrm{MXC}\right)$ then very well describes the experimental data in Fig.\,\ref{fig:exp1A}\,a (shown by dashed lines).

\begin{table*}[t!]
\caption{\label{tab:fitting_params} Parameters of the linear fits shown in Fig.\,\ref{fig:exp1B}\,c. 
}
\begin{ruledtabular}
\begin{tabular}{ccccc}
\multirow{2}{*}{Fitting Parameters} & \multicolumn{4}{c}{Sample} \\
 & R4-I & R2-I & R3-II & Q2-III \\ \hline
$d n_\mathrm{eff} / d n_\mathrm{MXC}$ & $0.849\pm 0.014$ & $0.812\pm 0.012$ & $0.596\pm 0.012$ & $0.673\pm 0.024$ \\
$n_\mathrm{eff}^0,\,\times 10^{-3}$ & $24.0\pm 0.7$ & $18.5\pm 0.7$ & $18.1\pm 0.6$ & $12.6\pm 1.1$ \\
\hline
$T_\mathrm{env}$,\,mK & $102.0\pm 1.6$ & $162.9\pm 3.4$ & $127.0\pm 1.7$ & $100.3\pm 0.7$ \\ \hline
$d \gamma_1 / d n_\mathrm{eff}$, $1/\mu s$& $-0.007\pm0.18$ & $0.81\pm0.25$ & $1.04\pm0.25$ & $0.37\pm0.40$ \\
$\gamma_1^0$, $1/\mu s$ & $0.19\pm0.04$ & $0.41\pm0.07$ & $0.32\pm0.06$ & $0.38\pm0.08$ \\
\end{tabular}
\end{ruledtabular}
\end{table*}

Similarly we can plot the relaxation rate $\gamma_1 = 1/\tau_1$ against $n_\mathrm{eff}$ to check if Eq.\,(\ref{eq:gamma1_vs_n_mxc}) works for our data (see Fig.\,\ref{fig:exp1B}\,b). We fit the linear region and extract $\gamma_1^0$ (see results in Tab.\,\ref{tab:fitting_params}) which will be later used to describe the temperature dependence of the relaxation time $\tau_1$. However, the fitting error here is significant and we can only roughly estimate the ratio between the observed slope and the offset $\left(d\gamma_1/d n_\mathrm{eff}\right)/ \gamma_1^0\simeq 2$. This value cannot be observed for sample R4-I, where the relaxation rate is constant, until quasiparticle relaxation appears (see discussion further in Section\,\ref{sec:temperature}).

\subsubsection{\label{sec:n_eff_high_T}High-\texorpdfstring{$T_\mathrm{MXC}$}{T_MXC} Behavior}
Let us find the asymptotic behavior of $T_\mathrm{eff}$ at high temperatures of the cryostat. We can use Eq.\,(\ref{eq:n_eff_2}) to write
\begin{equation}
    n_\mathrm{eff} = \frac{\gamma_\mathrm{MXC}}{\gamma_1^0}\,n_\mathrm{MXC}
    + n_{\mathrm{eff}}^0,
\end{equation}
where photon number $n_\mathrm{eff}^0= \gamma_\mathrm{env}/\gamma_1^0 n_\mathrm{env}$ originates from the effective environment.
Assuming that 
$n_\mathrm{eff}\gg n_\mathrm{eff}^0$ and in the high temperature regime $\left(\hbar\omega_{ge}/k_B T_\mathrm{MXC}\ll 1\right)$
\begin{equation}
    n_\mathrm{eff} \approx \frac{\gamma_\mathrm{MXC}}{\gamma_1^0} n_\mathrm{MXC}\approx\frac{\gamma_\mathrm{MXC}}{\gamma_1^0} \frac{k_B T_\mathrm{MXC}}{\hbar \omega_{ge}}.
\end{equation}
Now from Eq.\,(\ref{eq:n_eff}) we get
\begin{equation}
    T_\mathrm{eff} = \frac{\hbar \omega_{ge}}{k_B}\frac{1}{\ln{\left(1+\frac{1}{n_\mathrm{eff}}\right)}}\approx \frac{\hbar \omega_{ge}}{k_B}\frac{1}{\ln{\left(1+\frac{\gamma_1^0}{\gamma_\mathrm{MXC}}\frac{\hbar \omega_{ge}}{k_B T_\mathrm{MXC}}\right)}}.
\end{equation}
If the coupling to the cryostat is relatively strong, i.e. $\gamma_\mathrm{MXC}/\gamma_1^0\lesssim1$ and $\frac{\gamma_1^0}{\gamma_\mathrm{MXC}}\frac{\hbar \omega_{ge}}{k_B T_\mathrm{MXC}}\ll1$, we can Taylor expand the logarithm resulting in
\begin{equation}
    \label{eq:T_eff_T_mxc}
    T_\mathrm{eff} \approx \frac{\gamma_\mathrm{MXC}}{\gamma_1^0} T_\mathrm{MXC},
\end{equation}

\subsubsection{\label{sec:T_sat}Qubit Saturation Temperature}

Let us consider Eq.\,(\ref{eq:n_eff_2}) in the low cryostat temperature limit $n_\mathrm{eff}\rightarrow0$:
\begin{equation}
    T_\mathrm{eff} \approx \frac{\gamma_\mathrm{env}}{\gamma_1^0}\frac{k_B T_\mathrm{env}}{\hbar \omega_{ge}}.
\end{equation}
which gives
\begin{equation}
    \label{eq:T_sat}
    T_\mathrm{eff} = \frac{\hbar \omega_{ge}}{k_B}\frac{1}{\ln{\left[1+\frac{\gamma_1^0}{\gamma_\mathrm{env}}\left(\exp\frac{\hbar \omega_{ge}}{k_B T_\mathrm{env}}-1\right)\right]}}.
\end{equation}

Now let us consider the limit of the cold environment $\left(\hbar\omega_{ge}/k_B T_\mathrm{env}\gg 1\right)$:
\begin{equation}
    T_\mathrm{eff} \approx \frac{\hbar \omega_{ge}/k_B}{\ln{\left(1+\frac{\gamma_1^0}{\gamma_\mathrm{e}}\exp\frac{\hbar \omega_{ge}}{k_B T_\mathrm{env}}\right)}}\approx\frac{\hbar \omega_{ge}/k_B}{\ln{\frac{\gamma_1^0}{\gamma_\mathrm{env}} + \frac{\hbar \omega_{ge}}{k_B T_\mathrm{env}}}}
\end{equation}
which can be reformulated as
\begin{equation}
    \label{eq:T_sat_cold}
    \frac{1}{T_\mathrm{eff}} =  \frac{1}{T_\mathrm{env}} + \frac{1}{T_\gamma},
\end{equation}
where we defined $T_\gamma=\frac{\hbar\omega_{ge}}{k_B}\left(\ln{\frac{\gamma_1^0}{\gamma_\mathrm{env}}}\right)^{-1}$. One can interpret Eq.\,(\ref{eq:T_sat_cold}) in the following way: if the environment temperature is low, then lowering the effective qubit temperature can be done by decoupling from this bath. Moreover, decoupling should be exponentially big for linear decrease of $T_\mathrm{eff}$.

If the environment is hot, $\left(\hbar\omega_{ge}/k_B T_\mathrm{env}\ll 1\right)$, we can expand the exponent in Eq.\,(\ref{eq:T_sat}) and get
\begin{equation}
    T_\mathrm{eff} = \frac{\hbar \omega_{ge}}{k_B}\frac{1}{\ln{\left(1+\frac{\gamma_1^0}{\gamma_\mathrm{env}}\frac{\hbar \omega_{ge}}{k_B T_\mathrm{env}}\right)}}.
\end{equation}
In this limit we can consider the case of relatively strong coupling to the cryostat bath $\gamma_\mathrm{MXC}/\gamma_1^0\lesssim1$ and $\frac{\gamma_1^0}{\gamma_\mathrm{env}}\frac{\hbar \omega_{ge}}{k_B T_\mathrm{env}}\ll1$, which leads to linear dependence
\begin{equation}
    T_\mathrm{eff} = \frac{\gamma_\mathrm{env}}{\gamma_1^0}{}T_\mathrm{env}.
\end{equation}

The high temperature of the environment $T_\mathrm{env}=100-165$\,mK is most likely caused by improper shielding of the sample stage and filtering of the measurement lines \cite{Simbierowicz_2024}, nonequlibrium quasiparticles \cite{Jin_2015} or TLSes \cite{Kulikov_2020, Lisenfeld_2019}. Recent works demonstrated the possibility of achieving lower effective qubit temperatures of $20-45$\,mK \cite{Jin_2015, Scigliuzzo_2020, Somoroff_2023}. However in our work we do not try to distinguish concrete channels of qubit relaxation. In fact, all different mechanisms, ultimately leading to thermalization to the cryostat temperature are included in the cryostat heat bath, and all others are considered as the environment bath.

\subsection{\label{sec:numerics} Comparison of the Experimental Data with the Numerical Model of the Qubit Population Evolution}

In order to understand the deviation of the population functions $A,B$ and $C$ from the exact formulas in Tab.\,\ref{tab:ABC} and the qubit effective temperature $T_\mathrm{eff}$ from the cryostat temperature $T_\mathrm{MXC}$ at higher temperatures, we implement a numerical model which accounts for time evolution of the population distribution of a qubit. First, let us discuss important assumptions underlying the model, which are based on the experimental data. We assume, that the control sequences of $\pi$-pulses are short compared to the qubit dephasing and relaxation times $\left(3\Delta t_\pi < \tau_\varphi, \tau_1 \right)$. Then we are interested in tracking the time evolution of the population distribution, described by the diagonal terms of the density matrix. The duration of the qubit state readout $\Delta t_\mathrm{RO}$ is larger than or of the order of the relaxation time $\left(\Delta t_\mathrm{RO}\gtrsim \tau_1\right)$.
The model considers population of the three lowest states with finite decay times $\tau_1^{ge}$ and $\tau_1^{ef}$ of the $ge$- and $ef$-transitions, respectively. 

\begin{figure}[tb]
\includegraphics[width=0.48\textwidth]{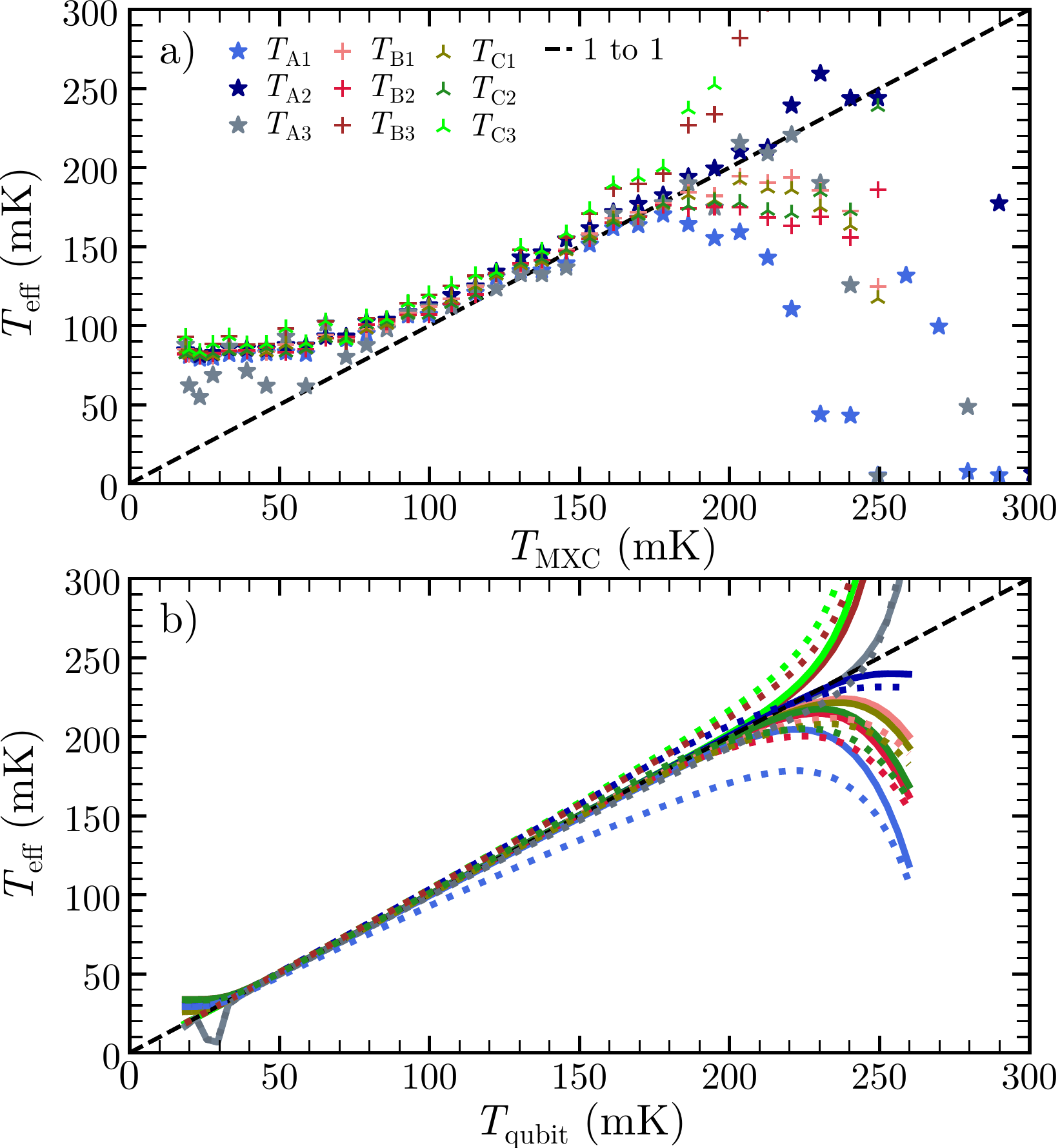}
\caption{\label{fig:model} 
Effective qubit temperature extracted from the population distribution measurements. 
a) Experimental data for the qubit R4-I for all nine ratios $A_1,A_2,A_3,B_1,\dots,C_3$. 
b) Effective qubit temperature, extracted from the qubit population distribution according to the numerical model (see details in main text). Using the device parameters of R4-I listed in Table\,\ref{tab:q_params}, we initialize the qubit population $p_i(T), i\in\{g,e,f\}$ Boltzmann-distributed at temperature $T_\mathrm{qubit}$ and imitate the measurement procedure. We consider the temperature dependence of qubit lifetime $\tau_1$, which we approximate from experimental data shown in Fig.\,\ref{fig:exp2}\,a. Solid lines show the result of simulating perfect $\pi$ pulses, which we compare to finite pulse efficiency $\delta_{ge}=0.9, \delta_{ef}=0.8$ (dotted lines). The simulation was done with parameters $2\tau_1^{ef}(T_\mathrm{qubit}\rightarrow 0) = \tau_1^{ge}(T_\mathrm{qubit}\rightarrow 0) = 5.5\,\mu s$, $\pi$-pulse duration is 165\,ns and readout pulse length $\Delta t_\mathrm{RO}=2\,\mu$s.
}
\end{figure}

We begin with a qubit, which is thermalized at a temperature $T$, setting initial populations $p_{g,e,f}^{(0)}(T)$ and then imitate application of $\pi_{ge}$- and $\pi_{ef}$-pulses by instantaneous swapping of the corresponding level populations (see Fig.\,\ref{fig:chip_pulses}\,e) with finite efficiencies, designated as $\delta_{ge,ef} \in [0,1]$. For example, a $\pi_{ge}$-pulse with $\delta_{ge}=0.9$ means that a qubit state with $p_g=1.0,\ p_e=0.0,\ p_f=0.0$ would turn result into $p_g=0.1,\ p_e=0.9,\ p_f=0.0$. Then we calculate the time evolution of the population during the readout time, and multiply it by arbitrary pure state responses $\varphi_i$. Next we construct the population ratios $A,\ B$ and $C$ and calculate the effective temperature (see Fig.\,\ref{fig:model} b). A detailed description of the model is provided in Appendix\,\ref{app:model}.

\begin{figure}[tb]
\includegraphics[width=0.48\textwidth]{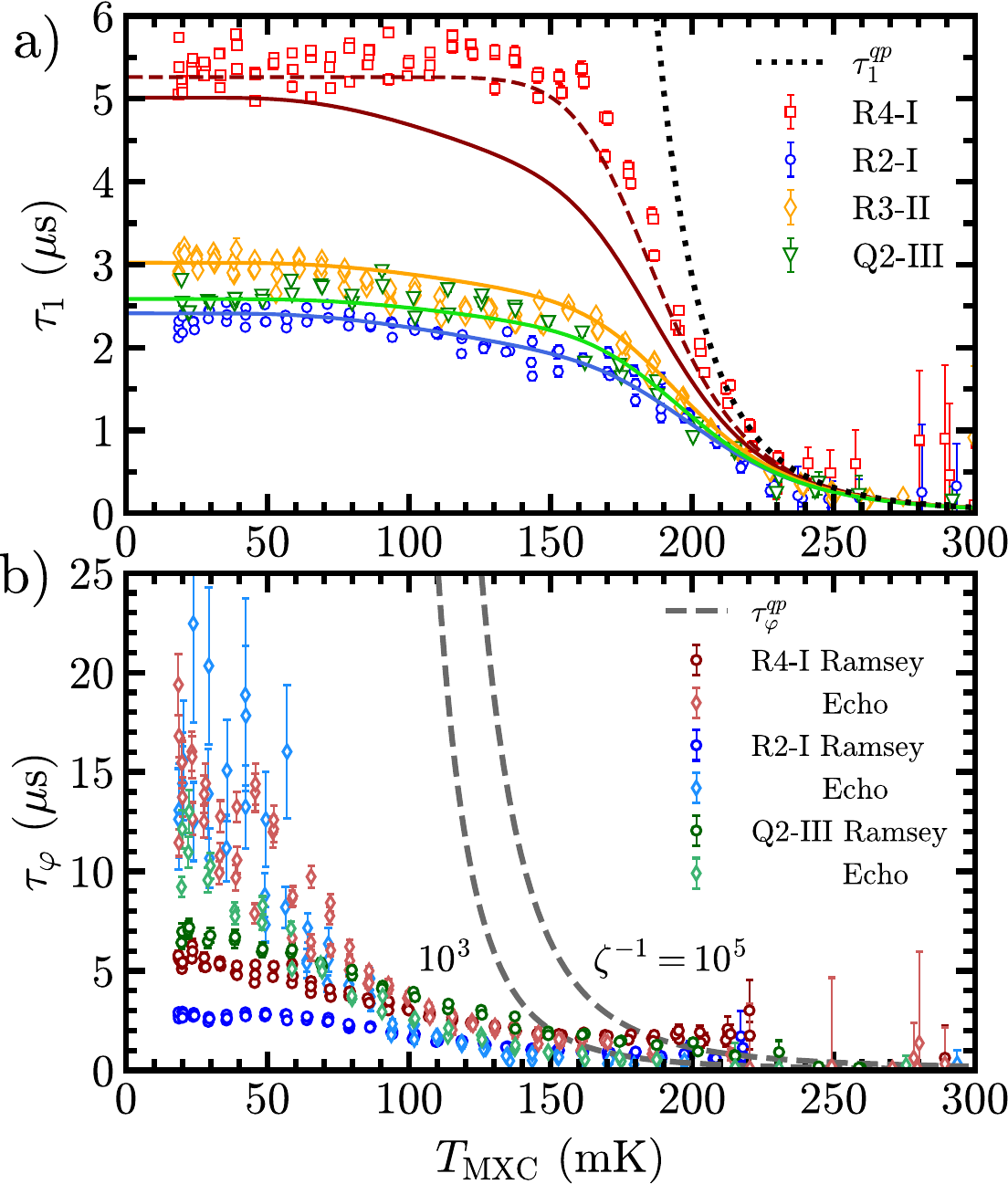}
\caption{\label{fig:exp2} 
a) Temperature dependence of the relaxation times $\tau_1$. Dotted black line shows suppression of $\tau_1^{qp}=1/\gamma_1^{qp}\left(T_\mathrm{MXC}\right)$ due to tunneling of equilibrium quasiparticles in Al, calculated using Eqs.\,(\ref{eq:gamma_qp}), the solid lines show the total relaxation time $\tau_1\left(T_\mathrm{MXC},T_\mathrm{eff}\right)$ (see Eq.\,(\ref{eq:tau_1_total})) and the dashed lines show the relaxation time as a sum of the quasiparticle-induced rate and the constant base-temperature rate, i.e. $\tau_1=1/\left[\gamma_1^{qp}\left(T_\mathrm{MXC}\right)+\gamma_1^{0}\right]$. b) Dephasing time $\tau_\varphi = 1/\gamma_\varphi$, extracted from measurements of $\tau_2^{R,E}$ and $\tau_1$, as a function of temperature. Symbols denote experimental data taken from Ramsey oscillations and Hahn echo measurements. Dashed lines corresponding to Eq.\,(\ref{eq:dephasing_qp}) show the increased dephasing rate due to the qubit frequency fluctuations caused by equilibrium quasiparticles.}
\end{figure}

The numerical model can qualitatively reproduce the deviation of the qubit population functions $A,B$ and $C$ from the analytical expressions (see Fig.\,\ref{fig:model_ABC} in Appendix\,\ref{app:model}), as well as the resulting deviations of the temperatures $T_\mathrm{eff}$ extracted with different methods in the higher temperature range, which is presented in Fig.\,\ref{fig:model}. This deviation is a consequence of the temperature-dependent qubit decay times $\tau_1^{ge}$ and $\tau_1^{ef}$, caused by quasiparticles. For the simulation results presented in Fig.\,\ref{fig:model}\,b, we use frequencies $\omega_{ge}$ and $\omega_{ef}$ and the temperature dependent relaxation time $\tau_1^{ge}(T)=1/\left[\gamma_1^{qp}(T)+\gamma_1^{0}\right]$, which correspond to sample R4-I shown as the red dashed line in Fig.\,\ref{fig:exp2}\,a, and $\tau_1^{ef}\approx\tau_1^{ge}/2$, valid for a transmon qubit \cite{Koch_2007}. For ideal $\pi$-pulses, the effective temperatures $T_{A1}, T_{A2}, \dots, T_{C3}$ extracted from different ratios $A_1,A_2,\dots,C_3$, respectively, demonstrate almost one-to-one behavior for the temperature range $30-200$\,mK, when the qubit lifetimes are relatively long. But above 200\,mK, when $\tau_1$ is significantly suppressed, there is an increasing deviation between these dependencies. Importantly, this does not indicate lack of thermalization of the qubit, rather than showing the limits of applicability of the given thermometry protocol. Moreover, in this temperature limit the values $T_{A1}, T_{A2}, \dots, T_{C3}$ start to depend on the pure state responses $\varphi_i$, which eliminates the main advantage of the protocol. In fact, Eqs.\,(\ref{eq:xy}) are no longer valid, as the qubit population changes significantly during the readout pulse.

Another error in determining the effective temperature of the qubit is the finite efficiency of the pulses. As seen in Fig.\,\ref{fig:model}\,b, the divergence of $T_{A1}, T_{A2}, \dots, T_{C3}$ grows monotonously with the temperature and is noticeable already at $100-150$\,mK, which is a characteristic of the protocol.  
However, for clarity in the estimates we deliberately introduce much coarser errors $\left(1-\delta_{ij}\sim0.1\right)$ than those existing in the experiment. For example, we can estimate, that population errors originating from imperfect $\pi$-pulses used in the measurements are much smaller, than those caused by other factors \cite{Steffen_2003, Motzoi_2009}. Thus, the population leakage to the $f$-level does not exceed $2\cdot10^{-6}$, compared to the steady-state population of $p_f\sim10^{-4}$ at 65\,mK.

\section{\label{sec:limits}Thermometer Limitations and Ways of Improvement}

\subsection{\label{sec:temperature}Temperature Range}

The Al-based transmon thermometers studied in this work showed a relatively narrow working temperature range $60-200$\,mK. In this subsection we discuss the limiting factors at higher temperatures. In this limit the following aspects should be taken into account:
i) consideration of only the three lowest energy states is valid until the population of the next third excited $d$-level is negligible, otherwise, at higher temperatures, the system (\ref{eq:xy}) should be supplemented with appropriate terms having $\varphi_d$ and $p_d$; ii) suppression of the qubit relaxation time $\tau_1$; iii) sufficiently long coherence time for the qubit state preparation.
We attribute the last two factors mainly to influence of equilibrium thermal quasiparticles appearing above 200\,mK.

\subsubsection{Suppression of Relaxation Time}

First, let us address the temperature dependence of the qubit decay time $\tau_1$, shown in Fig.\,\ref{fig:exp2}\,a. In the temperature range $20-150$\,mK it changes only slightly. 
As mentioned earlier, there is a noticeable difference between the low-temperature behavior of samples R2-I, R3-II and Q2-III following  Eq.\,(\ref{eq:gamma1_2baths}), and sample R4-I, whose relaxation rate appears to be temperature-independent below 170\,mK. This indicates the presence of other relaxation channels that are not described in our two-bath model.

Above $T_\mathrm{MXC}\approx170\,$mK a noticeable effect of quasiparticle-induced relaxation appears and already at $T_\mathrm{MXC}\approx250-270\,$mK decay time drops to below $0.5\,\mu$s, which is shorter than the readout pulse duration $\Delta t_\mathrm{RO}=1.5-2.0\,\mu$s and becomes comparable to the duration of $\pi$-pulse sequences used for the population measurements. To compare $\tau_1$ with theoretical prediction we use the quasiparticle relaxation model in Refs.\,\cite{Catelani_2011, Krause_2024} that gives the relaxation rate for a transmon as 
\begin{equation}
    \label{eq:gamma_qp}
    \begin{split}
        \gamma_1^{qp} = \frac{1}{\pi} & \frac{ \omega_p^2 }{ \omega_{ge} } 
        \left\{ x_{qp} \sqrt{ \frac{ 2\Delta }{ \hbar \omega_{ge} } } \right. \\
        & + \left. 4e^{-\frac{\Delta}{k_B T}} \cosh \left( \frac{\hbar\omega_{ge}}{2 k_B T} \right) \, K_0 \left( \frac{\hbar\omega_{ge}}{2 k_B T} \right) \right\} \\
    \end{split} 
\end{equation}

\noindent with the plasma frequency $\omega_p = \sqrt{8 E_J E_C}/\hbar$, equilibrium quasiparticle density $x_{qp} = \sqrt{2\pi k_B T/ \Delta}\,\exp\left(-\Delta/ k_B T\right)$, modified Bessel function of the second kind $K_0$ and the superconducting gap $\Delta$. Expression (\ref{eq:gamma_qp}) provides a good correspondence with the experimentally observed drop of $\tau_1$ at temperatures above 200\,mK for the gap value $\Delta_\mathrm{Al}/e = 180\,\mu\mathrm{V}$. Thus from Eqs.\,(\ref{eq:gamma1_2baths}) and (\ref{eq:gamma_qp}) we write the full expression for the temperature dependent relaxation time as 
\begin{equation}
    \label{eq:tau_1_total}
    \tau_1=1/\left\{\gamma_1^{qp}\left(T_\mathrm{MXC}\right)+\gamma_1^{0}\left[2n\left(\omega_{ge}, T_\mathrm{eff}\right)+1\right]\right\},
\end{equation}
which well agrees with the experimental data for samples R2-I, R3-II and Q2-III in Fig.\,\ref{fig:exp2}\,a.

In accordance with the numerical model, quasiparticle suppression of $\tau_1$ has a significant influence on the population measurements, which in turn adversely affects the determination of the effective temperature. This is a hard limit due to the exponential growth of the quasiparticle density with temperature. 

Another possibility of widening the dynamical range of the thermometer to higher temperatures is employment of materials with a larger superconducting gap to shift the quasiparticle suppression limit to higher temperatures. This might be done by replacing the tunnel junction with a nanowire made of granular Al having critical temperatures $T_c^\mathrm{grAl}$ up to 3.15\,K \cite{Grunhaupt_2019, Winkel_2020, Rieger_2023}, which is 2.5 times higher than the critical temperature of the bulk aluminium, $T_c^\mathrm{Al, bulk} = 1.2$\,K or by using Nb-based tunnel junctions \cite{Dmitiriev_1999, Dolata_2003, Anders_2009, Zheng_2008, Anferov_2024_2, Anferov_2024_1} with the critical temperature $T_c^\mathrm{Nb} = 7.6-9.3$\,K. Recently, all-Nb transmon qubits demonstrated $\tau_1\simeq 0.5 - 1.0\,\mu$s in the temperature range $0.5-1.0$\,K \cite{Anferov_2024_2, Anferov_2024_1}. 

\subsubsection{Suppression of Coherence}

The qubit state preparation is also of great importance in the population distribution measurement presented here. In this sense, the qubit dephasing time should be longer than the control pulse sequences $\left(\tau_\varphi \gtrsim 3 \Delta t_\pi\right)$. Since the dephasing rate is growing with the increase in temperature, it limits the duration of the pulse sequences at higher temperatures. 

The typical length $\Delta t_\pi$ of separate $\pi$-pulses varied between 60 and 220\,ns, with the lower limit $\Delta t_\mathrm{min}$ determined either by the qubit anharmonicity ($\Delta t_\mathrm{min}> h/|\alpha| \simeq 1/200\,$MHz $ = 5\,$ns) or by the dynamical range of the instruments and the total attenuation of the setup (see Fig.\,\ref{fig:setup}). We present experimental data of the temperature dependence of the dephasing time $\tau_\varphi$ in Fig.\,\ref{fig:exp2}\,b. The times $\tau_\varphi^R$ and $\tau_\varphi^E$ are extracted from the measurements of Ramsey oscillations and Hahn echo measurements, respectively, as $\tau_\varphi^{R,E} = 1/\gamma_\varphi^{R,E} = 1/\left(\gamma_2^{R,E} - \gamma_1/2\right)$, where the decoherence rates $\gamma_2^{R,E}$ describe Ramsey oscillations and Hahn echo experiments \cite{Krantz_2019, Blais_2021}. While in the low-temperature limit $\tau_\varphi^{R,E} \gtrapprox \tau_1$, with $\tau_\varphi^{R}$ exhibiting a clearly visible saturation below $50-70$\,mK, the dephasing time starts decreasing at lower temperatures than $\tau_1$ and reaches $1-2\,\mu$s at 150\,mK, becoming difficult to measure at $220-250\,$mK due to the finite ring-up time of the readout resonator $\sim Q_\mathrm{ext}/\omega_r\simeq 1\,\mu$s. Similarly to the analysis of the relaxation time, we can estimate the influence of quasiparticles on the dephasing, using the approach presented in Ref.\,\cite{Catelani_2012}. The pure dephasing rate
due to tunneling of equilibrium quasiparticles is given by expression
\begin{equation}
    \label{eq:pure_dephaging_qp}
    \gamma_\varphi = \frac{E_c }{\pi\hbar} \frac{k_B T}{\Delta} \exp{\left(-\frac{\Delta}{k_B T}\right)} ,
\end{equation}
and appears to be negligible in the experimental temperature range. However, we can estimate the dephasing mechanism caused by the Josephson frequency fluctuations due to quasiparticles occupying the Andreev bound states in the tunnel junction of the transmon. The latter writes \cite{Catelani_2012}
\begin{equation}
    \label{eq:dephasing_qp}
    \gamma_\varphi \sim 4 \pi \frac{ \omega_p^2}{\omega_{ge}}\sqrt{\frac{x_{qp}^A}{N_e}},
\end{equation}
where $x_{qp}^A = \exp{\left(-\Delta/k_B T\right)}$ is the equilibrium quasiparticle occupation of the Andreev states and $N_e = \zeta^{-1} g_T/2 g_K$ is the effective number of channels in the junction with the factor $\zeta^{-1}\sim 10^3 - 10^5$ describing transparency of the junction in the subgap regime \cite{Catelani_2011, Maisi_2011}, $g_T = 1/R_n$ being the junction conductance and $g_K = e^2/h$ the conductance quantum. In our case $R_n \approx 4.4-5.5\,$k$\Omega$, giving $g_T/2 g_K \approx 2.3-2.9$, and the dependence (\ref{eq:dephasing_qp}) for the realistic parameters is shown as dashed lines in Fig.\,\ref{fig:exp2}\,b.

The working temperature range is limited by the thermal quasiparticles in the basic operation as a cryostat thermometer. Yet when measuring objects on a chip coupled, e.g., by photon exchange, such a limitation becomes irrelevant, and the thermometer temperature range can be much wider (see further discussion in Section\,\ref{sec:meso_therm}).

\subsection{Signal-to-Noise Ratio and Accuracy}

\begin{figure}[tb]
\includegraphics[width=0.48\textwidth]{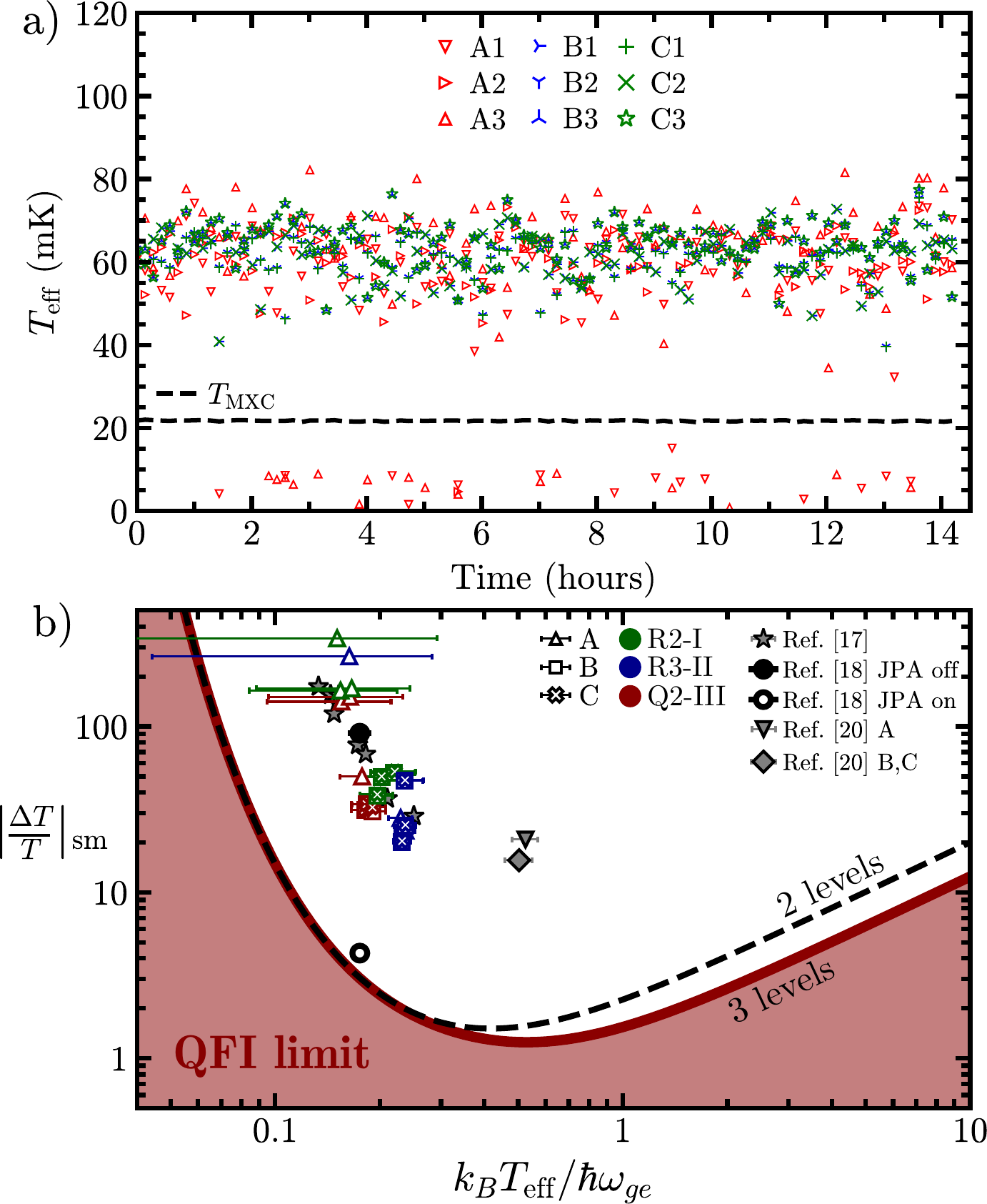}
\caption{\label{fig:meas_err} a) Time traces of the effective qubit temperature for sample Q2-III at the base cryostat temperature $T_\mathrm{MXC} \approx 22\,$mK (shown as the black dashed line).
b) Relative error of a single temperature measurement as a function of normalized effective temperature. The symbols denote experimental time averaged data obtained from time traces of three samples: R2-I (green), R3-II (blue) and Q2-III (red). The error bars show the standard deviation. 
We also show data from other works: Rabi oscillation measurements\,\cite{Jin_2015} (grey stars), correlations\,\cite{Kulikov_2020} (open and solid black circles) and original paper, where the $\pi$-pulse technique was described\,\cite{Sultanov_2021} (grey triangle and diamond).
The curves show the minimal achievable error for each experiment, which is defined by the Cramer-Rao bound and QFI, i.e. all data points should be above it. Solid red curve is the QFI limit obtained from Eq.\,(\ref{eq:QFI_3_levels}) for three-level case and the black dashed line depicts the two-level case Eq.\,(\ref{eq:QFI}). More information on the data shown in the figure can be found in Tab.\,\ref{tab:comparison}.
}
\end{figure}

Ultimately, the accuracy of the thermometer is limited by the quantum nature of the probe, which is reflected by the Cramer-Rao bound \cite{Frieden_1998}. This limit can be obtained from the quantum Fisher information (QFI) derived for a particular probe \cite{Mehboudi_2019}. For example, a two-level thermometer with energy separation $\hbar \omega_{ge}$ has the relative error of a single measurement $\left|\Delta T/\langle T\rangle\right|_\mathrm{sm}$ defined as
\begin{equation}
    \label{eq:QFI}
    \left(\frac{\Delta T}{\langle T\rangle}\right)^2_\mathrm{sm} \geq \frac{\left(1 + e^{x_{ge}}\right)^2}{x_{ge}^2 e^{x_{ge}}},
\end{equation}
$x_{ge}=\hbar\omega_{ge}/k_B T$.
Analogously an expression for a three-level anharmonic probe can be obtained as
\begin{equation}
    \label{eq:QFI_3_levels}
    \left(\frac{\Delta T}{\langle T\rangle}\right)^2_\mathrm{sm} \geq 
    \frac{\left(e^{x_{ge}+x_{gf}} + e^{x_{ge}} + e^{x_{gf}}\right)^2}{\left[x_{ge}^2 e^{x_{gf}} + x_{gf}^2 e^{x_{ge}} + \left(x_{ge} - x_{gf}\right)^2 \right] e^{x_{ge}+x_{gf}}},
\end{equation}
where $x_{gf}=\hbar\omega_{gf}/k_B T$ and $\omega_{gf} = \omega_{ge} + \omega_{ef}$.
Taking into account averaging over $N_\mathrm{av} = 2^{17}=131072$ repetitions, one gets the lower bound for the relative error $\left(\Delta T/\langle T\rangle\right)^2 = \left(\Delta T/\langle T\rangle\right)^2_\mathrm{sm}/N_\mathrm{av} \geq 5.0\cdot 10^{-5}$ or $\left|\Delta T/\langle T\rangle\right| \geq 0.7 \%$ for $\langle T\rangle = 65$\,mK and qubit frequency $\omega_{ge} = 7.04$ GHz. The noise equivalent temperature $\mathrm{NET}=\sqrt{\left|\Delta T\right|^2 \Delta t_\mathrm{meas}}$ with the measurement time $\Delta t_\mathrm{meas} = 29$\,s for our setup in the QFI-limited case would be $\mathrm{NET}_\mathrm{QFI}\approx 2.5$\,mK$/\sqrt{\mathrm{Hz}}$. 
However, in practice such values can hardly be achieved due to poor SNR of the measurement system, unless amplifiers with quantum limited noise are used \cite{Krantz_2019}.
In our case, the quantum limit for the temperature error is one order of magnitude lower than the experimental errors. For example, Fig.\,\ref{fig:meas_err}\,a represents a temperature time trace which was measured during several hours at the base cryostat temperature $T_\mathrm{MXC} = 22$\,mK. We calculate the average temperature and the deviation based on this statistics and recalculate it to the relative error per single measurement $\left|\Delta T/\langle T \rangle\right|_\mathrm{sm}=\sigma/\left(\langle T \rangle\sqrt{N_\mathrm{av}}\right)$.
The data have the relative errors $\left|\Delta T_A/\langle T_A \rangle\right| = 14-42 \%$, $\left|\Delta T_B/\langle T_B \rangle\right| = 8.6-9.4 \%$ and $\left|\Delta T_C/\langle T_C \rangle\right| = 8.7-9.4 \%$, which corresponds to $\mathrm{NET}\approx 28.5$\,mK$/\sqrt{\mathrm{Hz}}$ for $T_B$ and $T_C$.
It can be seen, that for each sample all data points have overlapping confidence intervals of $2\sigma$ (see Fig.\,\ref{fig:meas_err}). Typically, the temperatures extracted from the population ratio $A$ tend to have a worse accuracy and underestimate the qubit effective temperature, compared to $B$ and $C$, which give close values of the average temperatures and deviations. Thus, one can simply use the measurement with the minimal error to derive the temperature.

Accuracy of this thermometry technique can be compared with other works \cite{Sultanov_2021, Kulikov_2020}. In the latter a quantum limited JPA with SNR $\sim6$ was used for the measurements, which had shown a high relative accuracy of determining the excited state population ($\Delta p_e/p_e\approx 5\%$). In order to compare different measurement techniques, we calculated single measurement errors with respect to $N_\mathrm{av}$ for each experiment (which we take as standard deviation $\Delta T = \sigma$) and NET values for each experiment, using the data presented in the papers. We also argue, that the lowest effective qubit temperature shown in Ref.\,\cite{Kulikov_2020} was calculated correctly, and recalculate it from the excited state population using Eq.\,(\ref{eq:Teff}). Noteworthy, if the effective qubit temperature obtained in Ref.\,\cite{Kulikov_2020} had indeed matched the MXC stage temperature of 22\,mK, the accuracy per single measurement  would have significantly exceeded the Cramer-Rao limit. Results are shown in Fig.\,\ref{fig:meas_err}\,b and in Tab.\,\ref{tab:comparison}. Accuracy shown by the correlation measurements using a JPA almost reach the QFI limit, whereas other measurements appeared to be comparable to our results, which demonstrates the effectiveness of quantum limited amplifiers in improving the SNR.

\begin{table*}
\caption{\label{tab:comparison} 
Comparison with other experiments on qubit thermometry. For sample Q2-III three best measurements for $A$, $B$ and $C$ measurements are shown. Easiest comparison can be done by 
looking at error per single measurement $\left|\frac{\Delta T_\mathrm{eff}}{T_\mathrm{eff}}\right|_\mathrm{sm}$ or at the noise equivalent temperature (NET).}
\begin{ruledtabular}
\begin{tabular}{cccccccccccc}
 Sample ID & Measurement & $T_\mathrm{eff}$,\,mK & 
 $\omega_{ge}/2\pi$,\,GHz
 & $N_\mathrm{av}$ & $\left|\frac{\Delta T_\mathrm{eff}}{T_\mathrm{eff}}\right|_\mathrm{sm}$
 & $\Delta t_\mathrm{meas}$,\,s & NET, mK$/\sqrt{\mathrm{Hz}}$ & $T_\mathrm{MXC}$,\,mK\\ 
 \hline \\[-1em]
 Q2-III & $A2$ & $60.1\pm8.3$ & 7.04 & $2^{17}$ & 49.8 & 29 & 44.7 & 22 \\
 & $B1$ & $61.3\pm5.3$ & 7.04 & $2^{17}$ & 31.4 & 2.84 & .5 & 22 \\
 & $C1$ & $61.3\pm5.3$ & 7.04 & $2^{17}$ & 31.3 & 2.84 & .5 & 22 \\ 
 \hline \\[-1em]
 Ref.\,\cite{Jin_2015} & lowest $T_\mathrm{MXC}$ & $31.8\pm1.42$ & 4.97 & $1.5\cdot10^6$ & 20.8 & 10 & 43.2 & 15 \\
 & highest $T_\mathrm{MXC}$ & $59.8\pm0.44$ & 4.97 & $1.5\cdot10^6$ & 15.6 & 10 & 31.2 & 60 \\ 
 \hline \\[-1em]
 Ref.\,\cite{Kulikov_2020} & JPA off & $52.63\pm3.11$ & 6.26 & $2^{20}$ & 91.0 & 900 & 93.3 & 22 \\
 & JPA on & $52.63\pm0.15$ & 6.26 & $2^{20}$ & 4.30 & 900 & 4.45 & 22 \\ \hline \\[-1em]
 Ref.\,\cite{Sultanov_2021} & $A$ & $169.9\pm14.4$ & 6.74 & $6\cdot10^4$ & 20.8 & 9 & 43.2 & 13 \\
 & $B,C$ & $162.8\pm10.4$ & 6.74 & $6\cdot10^4$ & 15.6 & 9 & 31.2 & 13 \\
\end{tabular}
\end{ruledtabular}
\end{table*}

Notably, if a thermometer has a constant level separation, the relative error grows exponentially at $T < \hbar \omega_{ge}/2k_B$ 
(see Fig.\,\ref{fig:meas_err}), 
which is, demonstrated to some extent by experimental data from Ref.\,\cite{Jin_2015}.
Our system is a three-level quantum thermometer with a constant anharmonic spectrum (``sub-optimal quantum thermometers'' \cite{Mehboudi_2019}). Knowing the dependence of QFI on the spectrum of the system, it is possible to optimize it for a better performance in the experimentally realistic temperature range by choosing an appropriate level separation and anharmonicity. 
A qubit with a flux-tunable spectrum could provide higher precision for thermometry, by appropriate adjustment of the ratio $\hbar \omega_{ge}/k_B T_\mathrm{eff}$.

Other parameters of the experimental setup can also be fine-tuned to improve both the temperature measurement accuracy and the range of the cryostat temperatures, so that it reaches the quasiparticle limit. First of all, improvement of the $\pi$-pulse efficiency affects positively on the measurement accuracy. Secondly, as mentioned above, the short lifetime of the qubit is the main restriction of the method. We can divide the qubit state decay into two parts: i) decay during the control sequence and ii) decay during the readout. The first process leads to change of the qubit population distribution $p_i$ and can be associated to imperfect control pulses, leading to increased deviation between $T_{A1}, T_{A2},\dots,T_{C3}$. Decay during the readout pulse is defined mainly by the ring-up time of the resonator, and diminishes distinguishability between the qubit states (i.e. difference between $\varphi_i$ on the $I-Q$ plane), thus reducing the SNR, and also distorting the extracted effective temperature. The main method of improving the SNR, except for using a quantum limited amplifier, is optimizing the ratio $\chi/ \kappa$ for the readout resonator, where $\chi$ is the dispersive shift and $\kappa$ is the linewidth, as well as an appropriate readout frequency detuning for a given $\chi/ \kappa$, which directly affects the pure state responses $\varphi_i$ \cite{Bianchetti_2010}.

Another factor directly affecting the efficiency of the $\pi$-pulses is temperature dependent properties (specifically S21) of the measurement lines of the cryostat and CPWGs on the chip. In particular, it is known, that superconducting CPWG resonators have a temperature dependent Q-factors and fundamental frequencies \cite{Gao_2008, Lindstrom_2009, Pappas_2011, Sage_2011, Zmuidzinas_2012, Goetz_2016, McRae_2020}, which can lead to changes in amplitude of the qubit drive signal. Moreover, the qubit frequency even for the single-junction transmons can slightly vary with temperature (less than 50\,kHz in our case). All these factors can lead to an increased deviation of the $\pi$-pulse parameters if a calibration procedure is not applied at every cryostat temperature point.

\subsection{Operation Speed}
The presented method refers to the equilibrium type of quantum thermometry, meaning that the population measurement is performed after the qubit thermalization to the bath, which happens on time scales not shorter than the relaxation time $\tau_1$. In the presented experiment the time delay between the consecutive measurements, which is necessary for the qubit relaxation after the readout and thermalization of the qubit, was $\sim 10\,\tau_1$, which takes up most of the measurement time, and defines the speed of the response of the system. Pulse application and the qubit state readout can be limited to sub-microsecond range. Finally, the total measurement time is defined by the averaging number $t_\mathrm{tot} \sim N_\mathrm{av} \cdot 10\,\tau_1 \simeq 10^6 \tau_1$, with $N_\mathrm{av}\simeq 10^5$, meaning a time scale of seconds or minutes per point, depending on $\tau_1$.

\subsection{\label{sec:meso_therm}Perspective Applications - On-Chip Thermometry of Mesoscopic Structures}

Currently we do not consider qubit thermometers to be advantageous over well-established techniques such as calibrated thermometry of macroscopic objects. However, using them in applications similar to mesoscopic thermometers \cite{Giazotto_2006, Feshchenko_2013, Feshchenko_2015, Karimi_2018, Torresani_2013, Nicolí_2019}, such as NIS-junctions, Coulomb-blockade thermometers (CBT), shot noise thermometers (SNT) or thermometers based on quantum dots may be perspective, especially, taking into account qubit's minimal heat capacity (single photon needed for excitation) and low power readout (in single or few photons regime), which makes this thermometry highly noninvasive for the measured object.

In case of thermometry of a specific heat bath, coupled to a qubit, we can apply the same considerations as in Section\,\ref{sec:2_baths}. Let us assume, that the the effective environment is the object of our measurements. Then if the coupling in known, its temperature can be found from the saturation temperature of the qubit (see Eq.\,(\ref{eq:T_sat})). The latter is valid even in the case of the weak coupling and low temperatures of the object (Eq.\,(\ref{eq:T_sat_cold})). If the coupling to the object is unknown, but the bath temperature is variable, the dependence $T_\mathrm{eff}\left(T_\mathrm{env}\right)$ can still be measured. However, in reality the situation can be complicated by presence of other parasitic heat baths, which could reduce the sensitivity of the qubit to the temperature of the object.

An example of possible realization of qubit thermometry of another mesoscopic object is coupling a normal metal resistor with well defined properties such as resistance $R$ and temperature $T$ capacitively or inductively to a qubit \cite{Senior_2020, Cattaneo_2021, Pekola_2021, Gubaydullin_2022, Lemziakov_2024, Upadhyay_2024}. In this case, as shown in Eq.\,(\ref{eq:n_eff_2}), the qubit will be thermalized to the resistor if its relaxation rate to the resistor $\gamma_R$ will be much larger, than the total relaxation through all other channels, i.e. $\gamma_R \approx \gamma_1^0$, but low enough to allow for pulse measurements of the qubit, i.e. $1/\gamma_R \gtrsim 10\,\mu$s.

\section{\label{sec:conclusion}Conclusion}
In this work we experimentally realized equilibrium quantum thermometry of the cryostat temperature, based on the population distribution measurements of the three lowest energy levels of a transmon qubit. The presented technique allowed for temperature measurements in the range of $60-200\,$mK, in which the qubit population follows the Maxwell-Boltzmann distribution. 
Experimental data can be well explained by a two-bath model, which accounts for presence of an effective environment with a fixed finite temperature, different from that of the cryostat. We provided analysis of the effective temperature of the qubit, coupled to both cryostat and the environment having different temperatures and relative coupling strength.

We analyzed the working range of the qubit thermometer based on experimental data and numerical models and discussed possible ways to improve it. The temperature range of the thermometer was defined in the low temperature limit by the saturation temperature of $65-85\,$mK and above $200$\,mK by quasiparticle suppression of the qubit relaxation time. The experimental data were compared to a numerical model, describing time evolution of the qubit population, which showed good agreement. We briefly discuss a prospective application of the qubit thermometer for measurements of on-chip mesoscopic heat baths. Finally, we discussed the accuracy of the employed technique and compared it with other experiments and theoretical limits. We believe, that this experimental platform allows for exploitation of quantum thermometry algorithms including nonequilibrium techniques for faster measurements and dynamical control of the thermometers for enhanced precision.

\begin{acknowledgments}
The authors are grateful to Gershon Kurizki, Mohammad Mehboudi and Luis Alberto Correa Marichal for fruitful discussions and Yu-Cheng Chang for assistance with measurements and fabrication.
We acknowledge the provision of facilities and technical support by Aalto University at OtaNano - Micronova Nanofabrication Centre and OtaNano - Low Temperature Laboratory. 
\end{acknowledgments}

\appendix

\section{\label{app:single_bath}Relaxation and Excitation Rates for Coupling to a Thermal Bath}

Let us consider a quantum oscillator based on a resonator with frequency $\omega_0 = 1/\sqrt{LC}$ and characteristic impedance $Z_0=\sqrt{L/C}=50\,\Omega$, where $L$ and $C$ are the inductance and capacitance of the resonator, respectively, to which we connect an ohmic bath with resistance $R$ and temperature $T$. Then the excitation and relaxation rates $\Gamma_{\uparrow, \downarrow}$ caused by the bath can be written as \cite{Cattaneo_2021}
\begin{equation}
    \Gamma_{\uparrow,\downarrow} = \frac{Z_0}{R}\frac{\mp\omega_{0}}{1-\exp(\pm \hbar \omega_{0} \beta)},
\end{equation}

\noindent where $\beta = 1/k_B T$ is the inverse temperature and $k_B$ is the Boltzmann constant. For a more general case the resistance $R$ should be replaced by the real part of the bath impedance $Z$. Thus, we can introduce the quality factor $Q=\mathrm{Re}\, Z/Z_0$ and the resonator linewidth $\gamma = \omega_0/Q$, giving us
\begin{equation}
    \Gamma_{\uparrow,\downarrow} = \gamma\frac{\mp 1}{1-\exp(\pm \hbar \omega_{0} \beta)}.
\end{equation}
\noindent This can be further shortened for the sake of notation brevity to
\begin{eqnarray}
    &\Gamma_{\downarrow} = \gamma\left[ n\left(\omega, T\right)+1 \right],\\
    &\Gamma_{\uparrow} = \gamma\, n\,\left(\omega, T\right),
\end{eqnarray}
\noindent where $n=1/\left[\exp\left(\hbar\omega_0\beta\right) - 1\right]$ is the Bose-Einstein distribution.
The total decay rate in this case would be 
\begin{equation}
    \Gamma_1 = \Gamma_{\downarrow} + \Gamma_{\uparrow} = \gamma \left[ 2n\left(\omega, T\right)+1 \right].
\end{equation}

Analogous expressions can be readily applied to the qubit-bath interaction (see for example \cite{Karimi_2020, Cattaneo_2021}).

\subsection{Effective Temperature of a Two-Level System}
Let us consider a qubit with the ground state $\ket{g}$ and excited state $\ket{e}$, thermalized to a heat bath with temperature $T$. Then the population ratio of the qubit levels should follow the detailed balance principle $p_e/p_g=e^{-\hbar \omega \beta} = \Gamma_\uparrow/\Gamma_\downarrow$, and \cite{Theory_of_Open_Quantum-Systems}
\begin{equation}
    p_g = \frac{\Gamma_{\downarrow}}{\Gamma_{\downarrow}+\Gamma_{\uparrow}},\quad
    p_e = \frac{\Gamma_{\uparrow}}{\Gamma_{\downarrow}+\Gamma_{\uparrow}}.
\end{equation}
Also, we can write it in terms of the qubit polarization:
\begin{equation}
   \langle \sigma_z \rangle = -p_g + p_e = -\tanh{(\hbar \omega \beta/2)}.
\end{equation}

Energy relaxation rate is just a sum of the two rates:
\begin{equation}
    \gamma_1 = \frac{1}{\tau_1} = \Gamma_{\downarrow}+\Gamma_{\uparrow}.
\end{equation}

\subsection{Transmon Qubit Coupled to Several Baths}

Here we will extend the results from the previous section to the case when a quantum object interacts with a number of uncorrelated baths with different temperatures $T^{(i)}$ and coupling rates $\Gamma_{\uparrow,\downarrow}^{(i)}$. As in previous section, the only requirement for the baths is that the corresponding rates obey the detailed balance principle. In this case we can just sum up the rates when calculating the level population:
\begin{eqnarray}
\label{eq:pe_pg_many_baths}
    &p_g = \sum_{i} \Gamma_{\downarrow}^{(i)}/\sum_{i}\left(\Gamma_{\downarrow}^{(i)}+\Gamma_{\uparrow}^{(i)}\right),\nonumber\\
    &p_e = \sum_{i} \Gamma_{\uparrow}^{(i)}/ \sum_{i} \left(\Gamma_{\downarrow}^{(i)}+\Gamma_{\uparrow}^{(i)}\right),\nonumber\\
    &\frac{p_e}{p_g}=\frac{\Sigma_{i} \Gamma_{\uparrow}^{(i)}}{\Sigma_{i} \Gamma_{\downarrow}^{(i)}}=\frac{\Sigma_{i} \gamma^{(i)} n\left(T^{(i)}, \omega\right)}{\Sigma_{i} \gamma^{(i)} \left[n\left(T^{(i)}, \omega\right)+1\right]}
\end{eqnarray}

The qubit energy relaxation rate writes
\begin{equation}
\label{eq:gamma1_many_baths}
    \gamma_1 = \sum_{i} \left(\Gamma_{\downarrow}^{(i)}+\Gamma_{\uparrow}^{(i)}\right)=\sum_{i} \gamma^{(i)}\left[2\,n\left(\omega, T^{(i)}\right)+1\right].
\end{equation}

One can use the same considerations for a resonator (harmonic oscillator). Then again the detailed balance principle should be applied to get $p_{n+1}/p_n = \sum_{i} \Gamma_{\uparrow}^{(i)}/ \sum_{i} \Gamma_{\downarrow}^{(i)}$, which is positive, less then unity and does not depend on $n$. Thus we can find such a number $\beta_\mathrm{eff} = 1/k_B T_\mathrm{eff}$, to have $p_{n+1}/p_n = \exp{\left(-\hbar \omega\beta_\mathrm{eff} \right)}$, leading to
\begin{equation}
\label{eq:nq_2baths}
    \Bar{n} = \frac{\sum_{i} \Gamma_{\uparrow}^{(i)}}{\sum _{i}\left(\Gamma_{\downarrow}^{(i)}-\Gamma_{\uparrow}^{(i)}\right)} = \frac{\sum_{i} \gamma^{(i)} n\left(T^{(i)}, \omega\right)}{\sum_{i}\gamma^{(i)}}= n\left(T_\mathrm{eff}, \omega\right).
\end{equation}

Effective resonator temperature is then dependent on the number of photons $\Bar{n}$:
\begin{equation}
    T_\mathit{eff}^r = \frac{\hbar \omega_{r}}{k_B} \left(\ln{\frac{\Bar{n}+1}{\Bar{n}}}\right)^{-1}.
\end{equation}

\section{\label{app:model}Numerical Model}
To describe the dynamics of the qubit population distribution, firstly we neglect effects of the readout resonator, and secondly, restrict ourselves by considering only the three lowest energy states of the transmon, $\ket{g},\ \ket{e},\ \ket{f}$.
We assume that the initial qubit state is thermal, which means that its density matrix is diagonal. Then we track the dynamics of only these diagonal terms, characterizing the population distribution, which significantly simplifies the calculations.
We include only sequential decay and excitation processes with rates $\Gamma^{ge}_{\downarrow,\uparrow}$ and $\Gamma^{ef}_{\downarrow,\uparrow}$ for $g-e$ and $e-f$ transitions, respectively. Then the free time evolution of the qubit populations can be described by a system of differential equations:
\begin{align}
\label{eq:pop_de}
    \begin{cases}
        \dot{p}_f &= p_e\Gamma^{ef}_\uparrow-p_f\Gamma^{ef}_\downarrow, \\
        \dot{p}_e &= p_g\Gamma^{ge}_\uparrow + p_f\Gamma^{ef}_\downarrow-p_e(\Gamma^{ge}_\downarrow + \Gamma^{ef}_\uparrow), \\
        \dot{p}_g &= p_e\Gamma^{ge}_\downarrow -p_g\Gamma^{ge}_\uparrow ,
    \end{cases}
\end{align}
where dotted functions denote time derivatives. The system has an analytical solution:
\begin{align}
\label{eq:pop_solution}
    p_f(t) &= \zeta_f e^{\alpha_0 t} + \eta_f e^{\alpha_1 t} + \xi_f, \\
    p_e(t) &= \zeta_e e^{\alpha_0 t} + \eta_e e^{\alpha_1 t} + \xi_e, \\
    p_g(t) &= \zeta_g e^{\alpha_0 t} + \eta_g e^{\alpha_1 t} + \xi_g.
\end{align}

We find the following coefficients:
\begin{equation}
\begin{aligned}
    \alpha_{0,1} = \frac{1}{2}&\left\{-\left(\Gamma^{ef}_\uparrow + \Gamma^{ef}_\downarrow + \Gamma^{ge}_\uparrow + \Gamma^{ge}_\downarrow\right)\right. \\
    &\pm \left[\left(\Gamma^{ef}_\uparrow + \Gamma^{ef}_\downarrow + \Gamma^{ge}_\uparrow + \Gamma^{ge}_\downarrow\right)^2 \right. \\
    &- \left.\left. 4\left(\Gamma^{ge}_\downarrow\Gamma^{ef}_\downarrow + \Gamma^{ge}_\uparrow\Gamma^{ef}_\uparrow + \Gamma^{ge}_\uparrow\Gamma^{ef}_\downarrow\right)\right]^\frac{1}{2}\right\},
\end{aligned}
\end{equation}
\begin{align}
    \zeta_f &= \frac{\Gamma^{ef}_\uparrow\left(p_e^{(0)} - \xi_e\right) - \left(\Gamma^{ef}_\downarrow + \alpha_1\right)\left(p_f^{(0)}-\xi_f\right)}{\alpha_0-\alpha_1}, \\
    \zeta_e &= \frac{\alpha_0+\Gamma^{ef}_\downarrow}{\Gamma^{ef}_\uparrow}\zeta_f, \hspace{0.5cm} \zeta_g = \frac{\Gamma^{ge}_\downarrow}{\Gamma^{ef}_\uparrow}\frac{\alpha_0+\Gamma^{ef}_\downarrow}{\alpha_0+\Gamma^{ge}_\uparrow}\zeta_f, \\
    \eta_f &= \frac{\left(\Gamma^{ef}_\downarrow + \alpha_0\right)\left(p_f^{(0)}-\xi_f\right) - \Gamma^{ef}_\uparrow\left(p_e^{(0)} - \xi_e\right)}{\alpha_0-\alpha_1}, \\
    \eta_e &= \frac{\alpha_1+\Gamma^{ef}_\downarrow}{\Gamma^{ef}_\uparrow}\eta_f, \hspace{0.5cm} \eta_g = \frac{\Gamma^{ge}_\downarrow}{\Gamma^{ef}_\uparrow}\frac{\alpha_1+\Gamma^{ef}_\downarrow}{\alpha_1+\Gamma^{ge}_\uparrow}\eta_f,
\end{align}
\begin{align}
    \xi_f = \frac{1}{\mathcal{Z}}\frac{\Gamma^{ef}_\uparrow\Gamma^{ge}_\uparrow}{\Gamma^{ef}_\downarrow\Gamma^{ge}_\downarrow}, \hspace{0.5cm}
    \xi_e = \frac{1}{\mathcal{Z}}\frac{\Gamma^{ge}_\uparrow}{\Gamma^{ge}_\downarrow}, \hspace{0.5cm}
    \xi_g = \frac{1}{\mathcal{Z}},
\end{align}
where the canonical partition function $\mathcal{Z}=1+\Gamma^{ge}_\uparrow/\Gamma^{ge}_\downarrow + \Gamma^{ef}_\uparrow\Gamma^{ge}_\uparrow/\Gamma^{ef}_\downarrow\Gamma^{ge}_\downarrow$ imposes the normalization condition $p_g + p_e + p_f = 1$.

The schematic representation of the model is presented in Fig.\,\ref{fig:chip_pulses}\,c. We prepare the qubit with initial population $p_{g,e,f}^{(0)}$ and, for a fully thermalized qubit in the steady state ($t\rightarrow\infty$),  observe residual populations $p_{i}^{(\infty)} = \xi_{i}$. To mimic state manipulation with microwave pulses, we define 
pulse matrices acting on a population vector $\vec{p}(t) = (p_g(t), p_e(t), p_f(t))^T$:
\begin{align*}
    M_{ge} &= \begin{pmatrix}
        1-\delta_{ge} & \delta_{ge} & 0\\
        \delta_{ge} & 1-\delta_{ge} & 0 \\
        0 & 0 & 1
    \end{pmatrix},\\
    M_{ef} &= \begin{pmatrix}
        1 & 0 & 0\\
        0 & 1-\delta_{ef} & \delta_{ef} \\
        0 & \delta_{ef} & 1-\delta_{ef}
    \end{pmatrix},
\end{align*}
where we introduce a phenomenological efficiency parameter $0\leq\delta\leq 1$ to simulate effects such as variation of pulse length and amplitude, detuning and finite fidelity. 
Unlike the real control pulses, these matrices act on the qubit population instantaneously. It is a realistic assumption if the pulse lengths are much smaller than the relaxation times. In this case the $\pi$-pulse duration corresponds to the time delay between application of corresponding instantaneous population transformations.

\begin{figure}[tb]
\includegraphics[width=0.48\textwidth]{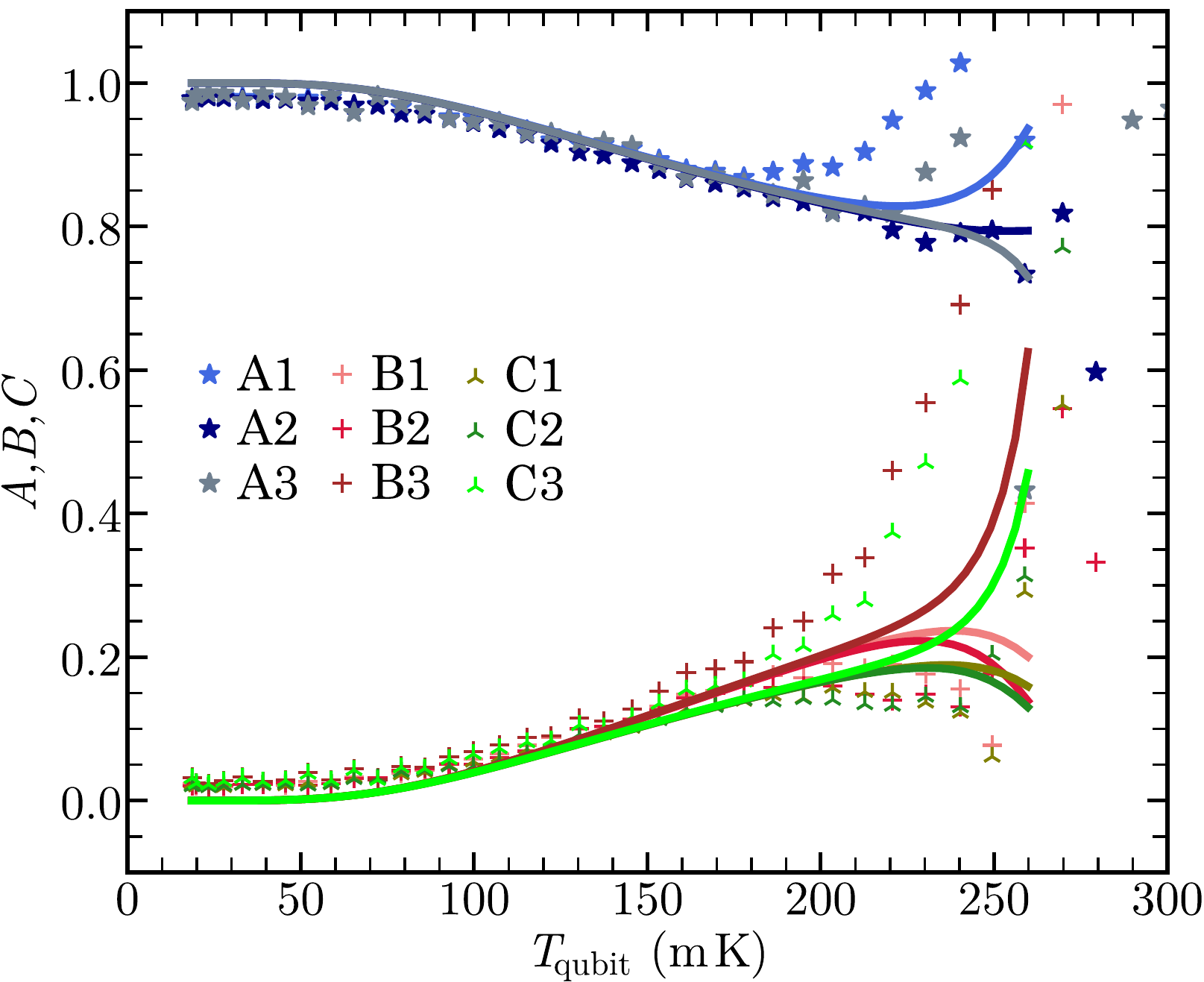}
\caption{\label{fig:model_ABC} 
Comparison between the qubit population measurements and the numerical simulation for qubit R4-I.
The simulation parameters are the same as in Fig.\,\ref{fig:model}\,b. The symbols show the experimental data and the lines correspond to the model.
} 
\end{figure}

\begin{figure}[tb]
\includegraphics[width=0.48\textwidth]{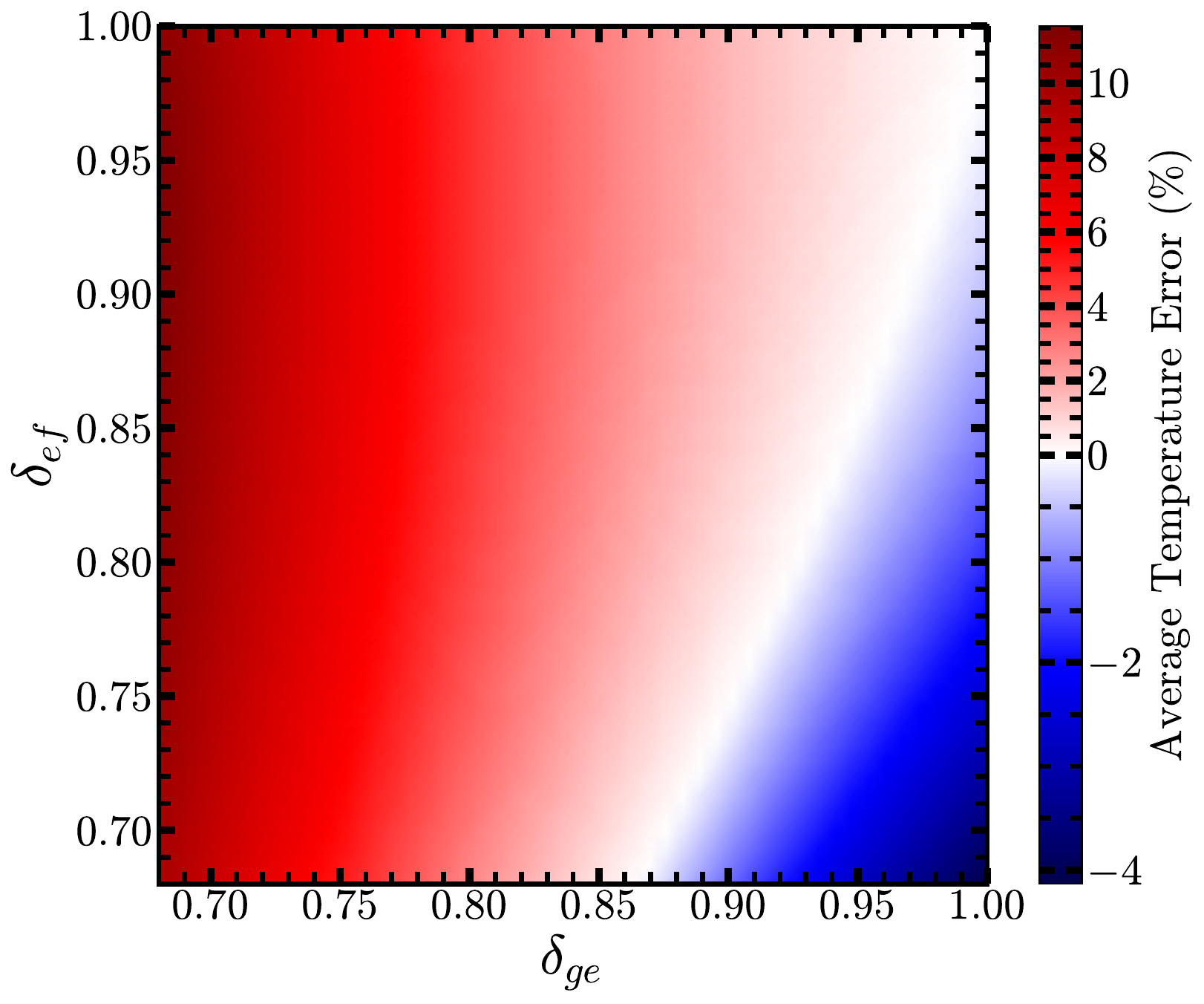}
\caption{\label{fig:model_efficiency} 
Averaged relative error of the effective temperature depending on the $\pi_{ge,ef}$-pulse efficiency, based on the numerical model at the qubit thermalization temperature 200\,mK. Apart from the variation in pulse efficiency, we use identical simulation parameters as in Fig.\,\ref{fig:model}. The relative error is calculated by comparing temperatures extracted from imperfect pulses with finite efficiencies $\delta_{ge}$, $\delta_{ef}$, denoted as $T_{\delta_{ge}, \delta_{ef}}$, and ideal pulses ($T_{\delta_{ge}=1, \delta_{ef}=1}$). We then average this quantity over all possible ways to extract qubit temperature $A_1, A_2, \dots, C_3$: $\langle [T_{\delta_{ge}, \delta_{ef}}(\alpha) - T_{\delta_{ge}=1, \delta_{ef}=1}(\alpha)] / T_{\delta_{ge}=1, \delta_{ef}=1}(\alpha)\rangle_{\alpha= A_1, A_2, \dots, C_3}$.
} 
\end{figure}

We studied influence of the pulse efficiency on the error of the qubit temperature measurement (see Fig.\,\ref{fig:model_efficiency}). Our model predicts the relative error less than 10\% at 200\,mK, which is comparable with the experimental data.

\section{\label{app:errors}Error Estimation}
\subsection{\label{app:error1}Mutual Dependencies}
As described in the main text, for thermometry we use temperature dependent functions $A, B$ and $C$ and each of them could be found in three different ways (see Table \ref{tab:ABC}), which gives nine different values for temperature. Though the calculation of these parameters based on six independent $x_{j},\ y_{k}$, only six of these functions $A_{i},\ B_{i},\ C_{i}$ are mutually independent. That could be easily shown from relation $A_i\times B_i=C_i$, $i \in \{1,2,3\}$. That also leads to the mutual dependencies for the corresponding measurement errors. Indeed, one can get the same relation between the relative errors of $A, B$ and $C$:
\begin{equation}
    \left(\frac{\Delta C}{C}\right)^2 = \left(\frac{\Delta A}{A}\right)^2 + \left(\frac{\Delta B}{B}\right)^2.
\end{equation}
From this equation one can get the relation between the absolute errors:
\begin{equation}
    (\Delta C)^2 = \left(\frac{C}{A}\right)^2 (\Delta A)^2 + \left(\frac{C}{B}\right)^2 (\Delta B)^2.
\end{equation}

In the low temperature limit we assume $p_f = 0$, $p_e\ll p_g \approx 1$. Using Eqs.\,(\ref{eq:A}), (\ref{eq:BC}) we can get $C/B \approx 1$ and $C/A \approx p_e$, and as a result $(\Delta C)^2 \approx (\Delta B)^2 + p_e^2 (\Delta A)^2$. If absolute errors for $A, B$ and $C$ are of the same order of magnitude (it is valid for most of the experimental results), absolute and relative errors for $B$ and $C$ are approximately equal: $(\Delta C)^2 \approx (\Delta B)^2$ and $|\Delta C/C| \approx |\Delta B/B|$.

In order to find the relation between the errors of $A, B$ and $C$ and the measured temperature $T$ one can use logarithmic derivatives of the functions $A(T), B(T)$ and $C(T)$. To simplify the final expressions, in the low temperature limit we can write
\begin{align}
    \label{eq:dT_A}
    &\left|\frac{\Delta T}{T}\right|_A = \frac{e^{x}}{x}|\Delta A| = \frac{e^{x}-1}{x}\left|\frac{\Delta A}{A}\right|, \\
    \label{eq:dT_B}
    &\left|\frac{\Delta T}{T}\right|_B = \frac{(e^{x}-1)^2}{x e^{x}}|\Delta B| = \frac{e^{x}-1}{x e^{x}}\left|\frac{\Delta B}{B}\right|, \\
    \label{eq:dT_C}
    &\left|\frac{\Delta T}{T}\right|_C = \frac{e^{x}}{x}|\Delta C| = \frac{1}{x}\left|\frac{\Delta C}{C}\right|,
\end{align}
where $x = \hbar \omega_{ge} / k_B T$. At low temperatures $x \rightarrow \infty$, Eqs.\,(\ref{eq:dT_B}) and (\ref{eq:dT_C}) give approximately equal errors, meaning that  the relative errors of the temperatures $T_B$ and $T_C$ are similar in the relevant temperature range. That agrees very well with the data presented in this paper (see Section\,\ref{sec:exp_details} and \ref{sec:limits}) and in Ref.\,\cite{Sultanov_2021}. 

Note, that although the coefficient between the relative error of $T$ and the relative error of $A$ in Eq.\,(\ref{eq:dT_A}) is $e^{\hbar \omega_{ge} / k_B T}$ times higher than the same coefficients in Eqs.\,(\ref{eq:dT_B}) and (\ref{eq:dT_C}), the $A$ itself is $e^{\hbar \omega_{ge} / k_B T}$ times higher than $B$ and $C$, so it means that relative error for $T_A$ could be the same as for $T_B$ and $T_C$. The exact relation between errors for $T_A$ and for $T_B$ and $T_C$ depends on particular pure state responses and voltage noise, as will be shown in the next subsection.

\begin{figure*}[t!]
\includegraphics[width=1.0\textwidth]{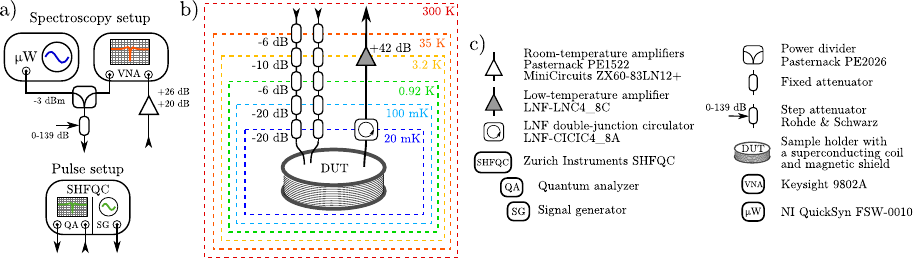}
\caption{\label{fig:setup} 
Experimental setup. a) Room-temperature part of the setup, used in the experiment. 
b) Cryostat setup used for the low-temperature measurements.
c) Decryption of the symbols, used in the panels a) and b), denoting instruments and other devices used in the measurement setup.
}
\end{figure*}

\subsection{\label{app:error2}Further Error Analysis}

One can notice that columns 1, 2 and 3 in Table \ref{tab:ABC} have the same structure: $A,\ B$ and $C$ from one column could be written in a form $A = b/c$, $B = a/b$ and $C = a/c$, where $a,\ b$ and $c$ are the measured voltage differences. For example, in the first column $a = x_2 - y_2$, $b = x_0 - x_1$ and $c = y_0 - y_1$. Also, one can write these voltage differences in the following form: $a = (p_e - p_f)\Delta \varphi_i$, $b = (p_g - p_e)\Delta \varphi_i$, $c = (p_g - p_f)\Delta \varphi_i$. Here $\Delta \varphi_i = \varphi_j - \varphi_k$ for $i,j,k \in\{g,e,f\}$, $i\neq j \neq k$. 

One can then make an additional assumption that the variance of a single measurement $x_i, y_i$ is a sum of the voltage noise variance (defined by the measurement setup) and the variance of protective measurements with operator $\hat{M}$ of the qubit in a thermal state with density matrix $\hat{\rho}_{th}$: $\sigma_{x,y}^2 = Tr(\hat{\rho}_{th} \hat{M}^2) - Tr(\hat{\rho}_{th} \hat{M})^2$. Based on this, one can find absolute errors for measurement differences $a, b, c$:
\begin{align}
    (\Delta a)^2 & =\ 2 p_e p_f \Delta \varphi_i^2\nonumber \\
    & + (p_e+p_f)p_g(\Delta \varphi_j^2 + \Delta \varphi_k^2) + \sigma_a^2, \\
    (\Delta b)^2 & =\ 2 p_e p_g \Delta \varphi_i^2\nonumber \\
    & + (p_e+p_g)p_f(\Delta \varphi_j^2 + \Delta \varphi_k^2) + \sigma_b^2, \\
    (\Delta c)^2 & =\ (p_g+p_e) p_f \Delta \varphi_i^2\nonumber \\
    &+ (p_g+p_f)p_e(\Delta \varphi_j^2 + \Delta \varphi_k^2) + \sigma_b^2,
\end{align}
here $\sigma_a^2, \sigma_b^2, \sigma_c^2$ are the measurement voltage variances, which in general are not the same for different parameters $a, b, c$ and could be found from the measurements. Relative errors for the voltage differences $a, b$ and $c$ could be written as:
\begin{align}
    \label{eq:da}
    \left(\frac{\Delta a}{a}\right)^2 = \frac{2 p_e p_f}{(p_e-p_f)^2} & + \frac{(p_e+p_f)p_g}{(p_e-p_f)^2} F(\Delta \varphi_i)\nonumber\\
    + & \frac{\sigma_a^2}{\Delta \varphi_i^2 (p_e-p_f)^2}, \\
    \label{eq:db}
    \left(\frac{\Delta b}{b}\right)^2 = \frac{2 p_e p_g}{(p_g-p_e)^2} & + \frac{(p_g+p_e)p_f}{(p_g-p_e)^2} F(\Delta \varphi_i)\nonumber\\
    + & \frac{\sigma_b^2}{\Delta \varphi_i^2 (p_g-p_e)^2}, \\
    \label{eq:dc}
    \left(\frac{\Delta c}{c}\right)^2 = \frac{(p_e+p_g) p_f}{(p_g-p_f)^2} & + \frac{(p_g+p_f)p_e}{(p_g-p_f)^2} F(\Delta \varphi_i)\nonumber\\
    + & \frac{\sigma_c^2}{\Delta \varphi_i^2 (p_g-p_f)^2}, 
\end{align}
where $F(\Delta \varphi_i) = (\Delta \varphi_j^2 + \Delta \varphi_k^2)/\Delta \varphi_i^2$, $i\neq j \neq k$. Note, that function $F(\Delta \varphi_i)$ reaches the minimum value of 0.5 when $\Delta \varphi_i = 2\Delta \varphi_j = 2\Delta \varphi_k,\ i\neq j\neq k$. This gives a condition for the optimal difference of the pure state responses. Importantly, this condition can be satisfied only for the one out of three differences $\Delta \varphi_i$ at the same time. 

In order to find relative errors for the $A, B, C$, one can use the following expressions:
\begin{align}
    \left(\frac{\Delta A}{A}\right)^2 = \left(\frac{\Delta b}{b}\right)^2 + \left(\frac{\Delta c}{c}\right)^2, \\
    \left(\frac{\Delta B}{B}\right)^2 = \left(\frac{\Delta a}{a}\right)^2 + \left(\frac{\Delta b}{b}\right)^2, \\
    \left(\frac{\Delta C}{C}\right)^2 = \left(\frac{\Delta a}{a}\right)^2 + \left(\frac{\Delta c}{c}\right)^2.
\end{align}
Using Eqs.\,(\ref{eq:da}, \ref{eq:db}, \ref{eq:dc}) for relative errors for $a, b$ and $c$, one can substitute it to the equations above and get the final (but extremely bulky) result.

As in a previous subsection, one can find absolute errors for $A, B, C$ in the low-temperature limit ($p_f = 0$, $p_e\ll p_g \approx 1$):
\begin{align}
    &(\Delta A)^2 \approx 2 p_e + F(\Delta \varphi_i) + \frac{\sigma_b^2}{\Delta \varphi_i^2} + \frac{\sigma_c^2}{\Delta \varphi_i^2}, \\
    &(\Delta B)^2 \approx p_e F(\Delta \varphi_i) + \frac{\sigma_a^2}{\Delta \varphi_i^2}, \\
    &(\Delta C)^2 \approx p_e F(\Delta \varphi_i) + \frac{\sigma_a^2}{\Delta \varphi_i^2}.
\end{align}
These equations illustrate several important features of errors of the functions $A, B$ and $C$. Firstly, the relative errors $\left|\Delta A/\langle A \rangle \right|, \left|\Delta B/\langle B \rangle \right|$ and $\left|\Delta C/\langle C \rangle \right|$ grow as $1/\sqrt{p_e} \propto \exp \left(\hbar \omega_{ge}/2k_B T\right)$ when $T \rightarrow 0$. Secondly, precision of the method could be limited by voltage noise, as it is for the measurements presented in this paper. Thirdly, in general, measurement precision highly depends on the differences $\Delta \varphi_i$ of the pure state responses. On one hand, it gives some room for the measurement protocol optimization in terms of maximization of the $\Delta \varphi_i^2$ and minimization of $F(\Delta \varphi_i)$. But, unfortunately, this optimization should be done for each qubit separately. Finally, one can obtain the minimum possible values for the absolute errors in the low temperature limit: $(\Delta A)^2 \approx 2 p_e +0.5$,  $(\Delta B)^2 \approx 0.5 p_e $, $(\Delta C)^2 \approx 0.5 p_e $. 

To sum up, the precision of errors for the temperature estimations from $A, B$ and $C$ measurements are limited by voltage noise and strongly depend on the differences $\Delta \varphi_i$ of the pure state responses. So, in general, out of nine possible ways to extract temperature, there is no unique or universal answer as to which one is the optimal one. Instead, for every particular sample and measurement setup one needs to compare them and choose the best one.

\section{\label{app:fab}Sample Fabrication}

The fabrication was done by a multi-step electron-beam lithography (EBL) in a 100\,kV EBL system Raith EBPG 5200.
First, pristine high-resistive ($\rho > 10^4$\,Ohm$\cdot$cm) undoped Si $\langle100\rangle$ 200\,mm wafers were RCA-cleaned and magnetron sputtered with 10\,nm of Al and 100\,nm of Nb without a vacuum break. Then the wafers were spincoated with 400\,nm thick layer of AR-P 6200.13 e-beam resist by spinning at 4000\,rpm for 60\,s and then baked at 160$^\circ$C for 9\,min. After that the samples were cleaved into $\sim$35$\times$35\,mm$^2$ pieces and loaded into the EBL system. Each wafer piece was used for fabrication of nine $7\times 7\,\mathrm{mm}^2$ chips. The exposure of the ground plane was done with the dose 350\,$\mu$C/cm$^2$, 200\,nA beam current and a 50\,nm step size. 
After that the e-beam resist was developed in AR-P 546-600 for 3\,min 30\,s and rinsed with IPA for 3\,min. The etching of Nb was done in a RIE tool Oxford Instruments Plasmalab 80 Plus. Prior to the etching the empty chamber with a dummy quartz wafer was precleaned with a CF$_4 + $O$_2$ process using 100\,sccm and 15\,sccm flows of CF$_4$ and O$_2$ respectively at the pressure 600\,mTorr, bias voltage 200\,V and RF power 200\,W during 10\,min. Then the samples were loaded into the chamber and pumped to 1$\cdot$10$^{-5}$\,mbar. The RIE etching was done with a SF$_6 +$Ar process at 20\,sccm flow of SF$_6$ and 10\,sccm flow of Ar at the pressure of 15\,mTorr, $V_\mathrm{bias}=360$\,V and RF power 100\,W during 90\,s. After that residues of the resist were removed with an O$_2$ descum process at the 40\,sccm flow, pressure 250\,mTorr, $V_\mathrm{bias}=320$\,V and RF power $150$\,W.
Then the samples were put into acetone for 4\,min, rinsed with IPA and dried with N$_2$ gun. Removal of the Al stopping layer was done using AZ351B developer during 2\,min. Then the samples were thoroughly rinsed with deionised water (DIW) for 10\,min and dried by putting in IPA for 5\,min with consecutive N$_2$ drying.

The next step was forming of Josephson junctions (JJs) (see Fig.\,\ref{fig:chip_pulses}\,f). For that the samples was spincoated by a MMA/PMMA e-beam resist bilayer. A 800\,nm thick layer of MMA EL-11 was applied by two steps of spincoating at 4000\,rpm and baking at 160$^\circ$C for 2\,min. Then a 200\,nm layer of PMMA A4 was applied by spinning at 4000\,rpm and baking at 160$^\circ$C for 10\,min. 
Finally, an EBL step was performed to create a suspended mask for further JJ shadow deposition. The main and undercut doses for exposure of submicron structures was 950\,$\mu$C/cm$^2$ and 150\,$\mu$C/cm$^2$, respectively, with 1\,nA beam current and 4\,nm stepsize.
The development of the mask was done in a two-step process: (i)
development in MIBK:IPA 1:3 mixture for 25\,s and (ii) in methylglycol:methanol 2:1 mixture for 15\,s with a following rinsing in IPA for 30\,s.
After that the samples were loaded into an e-beam evaporator Plassys MEB700S2-III UHV for JJ deposition. First, the samples were pumped down to $5\cdot10^{-7}$\,mbar in the loadlock and an Ar milling was done to remove the resist residues at Ar pressure of $4\cdot10^{-4}$\,mbar with ion-beam gun (IBG) parameters of 60\,mA beam current and 200\,V beam voltage at angles $\pm 12^\circ$ for 30\,s at each angle. Then the substrate was moved into the oxidation chamber and pumped to $1\cdot10^{-7}$\,mbar. After that the chambers were connected and Ti gettering was performed at a rate of 0.5\,\AA/s for 5\,min. Then the first layer of 25\,nm thick Al was deposited at angle $-12^\circ$ at 1\,\AA/s rate at pressure below $5\cdot10^{-8}$\,mbar in the oxidation chamber, after which the evaporation chamber was closed and an oxidation was done at a pressure of 18\,mbar during 7\,min. The second layer of 45\,nm thick Al was deposited at $+12^\circ$ at 1\,\AA/s rate at a pressure below $1\cdot10^{-7}$\,mbar in the oxidation chamber. Finally, another oxidation was performed at pressure 20\,mbar during 10\,min.
The lift-off procedure was done in hot acetone at $52^\circ$C
for 3\,h, then the samples were rinsed with IPA and dried with a N$_2$ gun.

Now the third step of EBL was done to create patches (see Fig.\,\ref{fig:chip_pulses}\,f) for a galvanic connection between JJs and the Nb ground plane. The MMA/PMMA e-beam resist bilayer was applied as described above and after that the samples were loaded into the EBL tool. The patch exposure was done with the dose of 950\,$\mu$C/cm$^2$ with 30\,nA beam current and 20\,nm stepsize. 
The resist was developed by putting into MIBK:IPA 1:3 mixture for 30\,s and then into methylglycol:methanol 2:1 for 30\,s followed by rinsing in IPA for 30\,s.
For the patch deposition we again used the Plassys tool. Analogously to the previous step, the samples in the loadlock were pumped to $5\cdot10^{-7}$\,mbar and a stronger Ar milling at zero angle was done for 2\,min to remove the surface Nb and Al oxides. The Ar pressure was $4\cdot10^{-4}$\,mbar and the IBG parameters were the following: 120\,mA beam current, 400\,V beam voltage.
Then the samples were moved to the oxidation chamber and pumped to $1\cdot10^{-7}$\,mbar, after which a deposition of 150\,nm of Al was performed at the rate of 2\,\AA/s and pressure below $1\cdot10^{-7}$\,mbar in the oxidation chamber with consecutive oxidation at 20\,mbar during 10\,min.
Another lift-off step was done similarly to the one described above.

For the low-temperature measurements the wafers were scribed with a diamond tip and cleaved into separate chips. The chips were mounted on a low-temperature sample holder using BF-6 glue and then wire-bonded using FS Bondtec 5330 with Al wire.


\bibliography{refs}
\end{document}